\documentclass[aps,pra,twocolumn,superscriptaddress,floatfix]{revtex4-2}
\usepackage{graphicx}
\usepackage{siunitx}

\usepackage{hyperref}
\usepackage[english]{babel}
\usepackage{braket}
\usepackage{nicefrac}
\usepackage{amsmath}
\usepackage{graphicx}
\usepackage{siunitx}
\usepackage{booktabs}
\usepackage{enumerate}
\usepackage{upgreek}
\usepackage{multirow}

\usepackage[rgb]{xcolor}
\definecolor{wierdred}{RGB}{170,0,68}
\definecolor{darkred}{RGB}{168,0,0}
\definecolor{darkblue}{RGB}{0,0,168}
\definecolor{lightred}{RGB}{255,0,102}
\definecolor{lightblue}{RGB}{0,204,255}
\definecolor{lightgreen}{RGB}{29,141,119}

\graphicspath{{./images/}}

\usepackage[rgb]{xcolor} 
\newcommand{\lrrm}[9]{$\mathrm{#1}(#2\,^2\mathrm{#3}_{#4})\mathrm{#5}(#6\,^2\mathrm{#7}_{#8}(F=#9))$}

\newcommand{\lrrmnohf}[8]{$\mathrm{#1}(#2\,^2\mathrm{#3}_{#4})\mathrm{#5}(#6\,^2\mathrm{#7}_{#8})$}
\newcommand{\ars}[4]{$\mathrm{#1}(#2\,^2\mathrm{#3}_{#4})$}
\newcommand{\arshf}[5]{$\mathrm{#1}(#2\,^2\mathrm{#3}_{#4}(F=#5))$}

\addto\extrasenglish{}

\addto\extrasenglish{}

\addto\extrasenglish{}

\addto\extrasenglish{}

\addto\extrasenglish{%
	
}

\def\equationautorefname#1#2\null{Eq.#1(#2\null)}

\newcommand{\new}[1]{\textcolor{black}{#1}}

\newcommand{\potassium}{$^{39}$K}
\newcommand{\caesium}{$^{133}$Cs}

\newcommand{\overbar}[1]{\mkern 1.5mu\overline{\mkern-1.5mu#1\mkern-1.5mu}\mkern 1.5mu}

\newcommand{\KCs}{$\overbar{\mathrm{K}}$Cs}
\newcommand{\CsK}{$\overbar{\mathrm{Cs}}$K}

\usepackage{units}
\usepackage{graphicx}
\usepackage{amsmath}
\usepackage{float}
\usepackage{braket}
\usepackage{multirow}

\usepackage{upgreek}
\newcommand{\diff}[2]{\frac{\mathrm{d} #1}{\mathrm{d} #2}}
\newcommand{\difftwo}[2]{\frac{\mathrm{d}^2 #1}{\mathrm{d} #2^2}}

\begin{document}


\title{Heteronuclear long-range Rydberg molecules}


\author{Michael Peper}
\affiliation{Laboratory of Physical Chemistry, ETH Z\"urich, 8093 Z\"urich, Switzerland}
\author{Johannes Deiglmayr}
\affiliation{Laboratory of Physical Chemistry, ETH Z\"urich, 8093 Z\"urich, Switzerland}
\affiliation{Department of Physics and Geoscience, University of Leipzig, 04109 Leipzig, Germany}
\email{johannes.deiglmayr@uni-leipzig.de}


\date{\today}

\begin{abstract}
We present the formation of homonuclear Cs$_2$, K$_2$, and heteronuclear CsK long-range Rydberg molecules in a dual-species magneto-optical trap for \potassium{} and \caesium{} by one-photon UV photoassociation. The different ground-state-density dependence of homo- and heteronuclear photoassociation rates and the detection of stable molecular ions resulting from auto-ionization provide an unambiguous assignment. We perform bound-bound millimeter-wave spectroscopy of long-range Rydberg molecules to access molecular states not accessible by one-photon photoassociation. Calculations based on the most recent theoretical model and atomic parameters do not reproduce the full set of data from homo- and heteronuclear long-range Rydberg molecules consistently. This shows that photoassociation and millimeter-wave spectroscopy of heteronuclear long-range Rydberg molecules provide a benchmark for the development of theoretical models.
\end{abstract}

\pacs{}

\maketitle

Long-range Rydberg molecules (LRMs) are bound states of a Rydberg atom and a ground-state atom located within the orbit of the Rydberg electron, where the binding results from the scattering of the almost free Rydberg electron off the ground-state atom~\cite{greeneCreationPolarNonpolar2000}. LRMs can have exotic properties~\cite{marcassa2014,shaffer2018ultracold}, such as kilo-Debye dipole moments \cite{li2011,booth2015} or bond lengths exceeding tens of nanometers, which have been exploited, \textit{e.g.}, to characterize anion shape resonances~\cite{engelPrecisionSpectroscopyNegativeIon2019}, pair-correlation functions~\cite{kinoshitaLocalPairCorrelations2005,whalenProbingNonlocalSpatial2019} and polaron dynamics~\cite{schmidtMesoscopicRydbergImpurity2016} in degenerate quantum gases. Whereas initial studies focused on homonuclear LRMs~\cite{bendkowsky09a,tallant2012,bellos2013,Anderson2014,niederprumGiantCrossSection2015,sassmannshausenprl2015,desalvo2015}, recently heteronuclear LRMs have attracted interest~\cite{eilesFormationLongrangeRydberg2018}. Proposed applications include the study of pair-correlation functions in atomic mixtures~\cite{eilesFormationLongrangeRydberg2018,whalenHeteronuclearRydbergMolecules2020}, and the creation of ultracold ion-pair systems with variable mass ratio~\cite{peperFormationUltracoldIon2020a,giannakeasDressedIonPairStates2020}.

In this letter, we employ heteronuclear LRMs to systematically study the properties of LRMs by isoelectronic substitution ~\cite{brownRotationalSpectroscopyDiatomic2003}. The electronic Hamiltonian of LRMs can be written as \cite{Anderson2014a,eilesHamiltonianInclusionSpin2017a}
\begin{equation}\label{eq:isoelectronic}
   H=H_{0}+H_\mathrm{SO}+H_\mathrm{HF} + V_\mathrm{FC},
\end{equation}
where $H_0$ is the Coulomb interaction of the ion core with the Rydberg electron, $H_\mathrm{SO}$ the spin-orbit interaction of the Rydberg electron, $H_\mathrm{HF}$ the hyperfine coupling between the electronic and nuclear spin of the ground-state atom, and $V_\mathrm{FC}$ the binding Fermi-contact interaction between the Rydberg electron and the ground-state atom. By substituting the atom in the Rydberg state with an atom of another species, only the contributions from $H_{0}$ and $H_\mathrm{SO}$ are altered. The spectra of initial and substituted LRMs should thus be reproduced with the same set of electron-atom scattering phases, allowing for a systematic test of theoretical models.

\begin{figure}
	\centering
	\includegraphics[width=\linewidth]{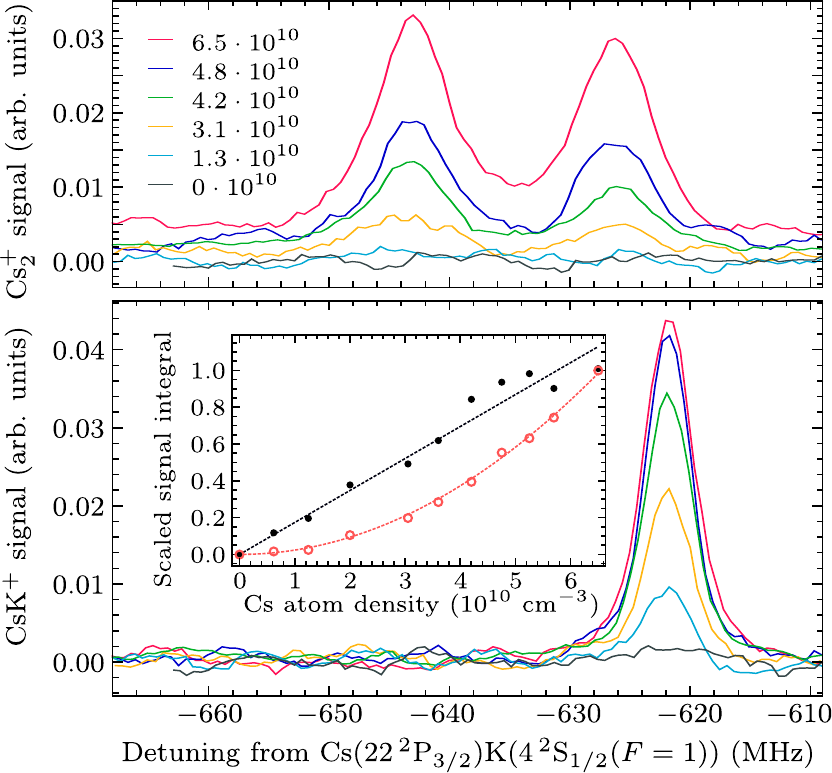}
	\caption{(upper panel) Cs$_2^+$ and (lower panel) CsK$^+$ ion signals as a function of the photoassociation laser detuning from the \ars{Cs}{22}{P}{3/2}$\leftarrow$\arshf{Cs}{6}{S}{1/2}{3} transition recorded after \SI{25}{\micro\second} of photoassociation. The Cs density was varied (legend in upper panel, values in cm$^{-3}$) and the K density was \SI{2.5(3)E10}{\per\centi\meter\cubed}. (inset) Full (open) circles are the scaled integrals of the CsK$^+$ (Cs$_2^+$) signal. Dashed lines represent linear and quadratic fits.
	\label{fig:density22}
	}
\end{figure}

\begin{figure*}[t]
	\centering
	\includegraphics[width=\textwidth]{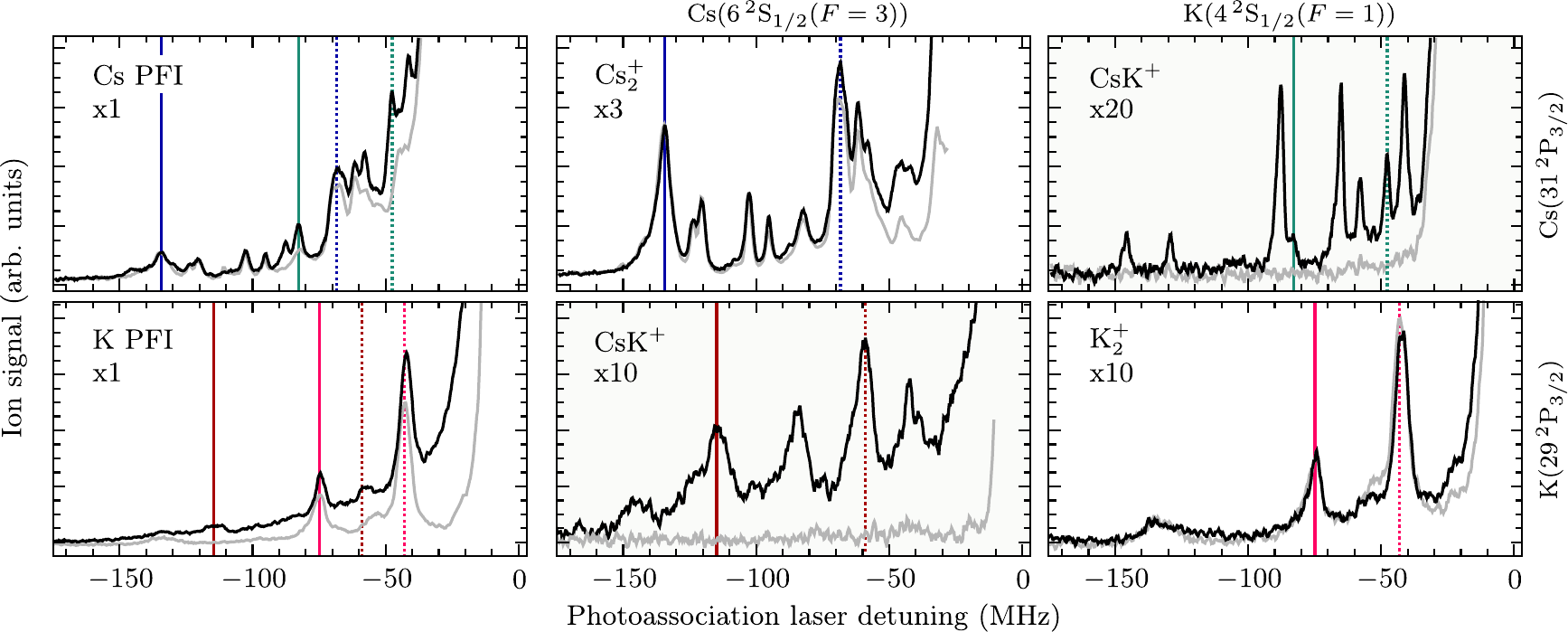}
	\caption{Photoassociation spectra of Cs (K) atoms prepared in the $F=3$ ($F=1$) ground state. (Upper row) Cs PFI (left), Cs$_2^+$ (middle), and CsK$^+$ (right) autoionization signal recorded below the \ars{Cs}{31}{P}{3/2}$\leftarrow$\arshf{Cs}{6}{S}{1/2}{3} transition. Resonances assigned to the $v=0$ level in the outermost well of the $^{3}\Sigma^+$ ($^{1,3}\Sigma^+$) states are marked by full (dotted) blue (Cs$_2$) and green (\CsK{}) lines. (Bottom row) K PFI (left), CsK$^+$ (middle), and K$_2^+$ (right) autoionization signal below the \ars{K}{29}{P}{3/2}$\leftarrow$\arshf{K}{4}{S}{1/2}{1} transition. Resonances assigned to the $v=0$ level in the outermost well of the $^{3}\Sigma^+$ ($^{1,3}\Sigma^+$) states are marked by full (dotted) magenta (K$_2$) and red (\KCs{}) lines. The gray dotted lines in the upper (lower) panels show the respective signals recorded in single-species Cs (K) MOTs. The signals have been averaged over 20 repetitions and a running average of 5, corresponding to a resolution of \SI{1.2}{MHz}. \label{fig:spectracomp3129}}
\end{figure*}

To this end, we perform spectroscopy of homo- and heteronuclear LRMs in a dual-species magneto-optical trap (dsMOT) of $^{133}$Cs and $^{39}$K atoms. By photoassociation below $n\,^2\mathrm{P}_{3/2}$ Rydberg states of cesium and potassium, we create homonuclear Cs$_2$ and K$_2$ as well as heteronuclear \KCs{} and \CsK{} LRMs, where the bar indicates the species in the Rydberg state. We probe the relative density distribution in the two-component gas mixture by mass-selective detection. Millimeter-wave spectroscopy (mmW) of \CsK{} LRMs yields additional data on $n\,^2\mathrm{S}_{1/2}$ Rydberg states.
A comparison to theoretical calculations reveals that current theoretical tools do not consistently describe the full set of our observations for homo- and heteronuclear LRMs below $n\,^2\mathrm{P}$ and $n\,^2\mathrm{S}$ asymptotes.

A dsMOT is formed by two background-vapour-loaded MOTs for $^{133}$Cs and $^{39}$K, spatially overlapped in an ultra-high vacuum chamber featuring electrodes for precise electric-field control and a micro-channel plate detector~\cite{deiglmayrpra2016,peperPrecisionMeasurementIonization2019}. The two atomic clouds are overlapped by coupling all laser beams for the two MOTs out of common single-mode fibers. The overlap is monitored by absorption images and is optimized by small changes to the alignment of the trapping beams and their intensities. Typical values for atom number, peak density, and temperature are \num{8e6} (\num{5e6}), \SI{2e10}{\per\cubic\cm} (\SI{4e10}{\per\cubic\cm}) and \SI{40}{\micro\kelvin} (\SI{20}{\micro\kelvin}) for the cesium (potassium) MOT. LRMs are formed by one-photon photoassociation using UV-laser pulses of \SI{10}{\micro\second} to \SI{30}{\micro\second} length and a peak intensity of \SI{200}{\watt\per\square\cm}. The UV radiation is obtained by frequency-doubling the output of a continuous-wave (cw) ring dye laser at 639 nm (571 nm) for Rydberg states of cesium (potassium). Its frequency is measured by a wavemeter (\textsc{HighFinesse} WS7), calibrated to the D2 line of potassium. Measurements with an optical frequency comb~\cite{deiglmayrpra2016} showed that this calibration procedure yields errors below \SI{1.5}{\mega \hertz} in the UV for frequency shifts of up to \SI{200}{\mega\hertz}. Following photoassociation, a ramped electric field is applied to state- and mass-selectively detect products~\cite{sassmannshausenprl2015}.

When photoassociation is performed in a mixture of species A and B, both homonuclear and heteronuclear molecules may be created. Pulsed-field ionization (PFI) does not allow us to distinguish between A$_2$ and $\overbar{\mathrm{A}}$B, because the molecule dissociates instantaneously upon removal of the Rydberg electron and in both cases $\rm A^+$ is detected. However, $\overbar{\mathrm{A}}$B (A$_2$) may autoionize to form a stable molecular ion $\rm AB^+$ ($\rm A_2^+$) \cite{sassmannshausenLongrangeRydbergMolecules2016,schlagmullerUltracoldChemicalReactions2016}. A photoassociated molecule is thus detected either as atomic or molecular ion depending on the autoionization rate, which varies strongly for different molecular states. \autoref{fig:density22} shows the Cs$_2^+$ and CsK$^+$ signals when the frequency of the photoassociation laser applied to the dsMOT is tuned below the \ars{Cs}{22}{P}{3/2}$\leftarrow$\arshf{Cs}{6}{S}{1/2}{3} transition. Clear resonances appear when the density of Cs ground-state atoms is increased. The detection of CsK$^+$ ions constitutes an unambiguous signature for the formation of heteronuclear LRMs. The number of heteronuclear ions depends linearly on the density of Cs ground-state atoms, whereas the number of Cs$_2^+$ ions exhibits a quadratic dependence, as expected for non-saturated photoassociation (see Appendix~\ref{appendix:padensities})~\cite{jones2006}.

\new{Having identified the autoionization products Cs$_2^+$, K$_2^+$, and CsK$^+$ as observables for the formation of homo- and heteronuclear LRMs, we systematically study the effects of isoelectronic substitution on the spectra of LRMs.} Photoassociation spectra below the Cs($31\,^2\mathrm{P}_{3/2}$) asymptote (upper row) and below the K($29\,^2\mathrm{P}_{3/2}$) asymptote (lower row) are shown in \autoref{fig:spectracomp3129}. \new{These asymptotes have been chosen because the effective quantum numbers $n^*=n-\delta$, where $\delta$ is the quantum defect, of the two states are almost identical (27.44~\cite{deiglmayrpra2016} and 27.29~\cite{peperPrecisionMeasurementIonization2019}, respectively)}. The photon energy required to excite K atoms into Rydberg states is sufficient to ionize Cs ground-state atoms. Because the cross section for this process is small, we observe only few Cs$^+$ ions when exciting K in the dsMOT.

The PFI signal obtained in the dsMOT (left panel, black line) exhibits dense series of resonances above $\SI{-150}{\mega\hertz}$, resulting from photoassociation of LRMs. The detuning of a resonance from the atomic Rydberg transition corresponds to the binding energy of the excited molecular level. At detunings above approx. $\SI{-30}{\mega\hertz}$, the molecular resonances are masked by off-resonant excitation of Rydberg atoms and Rydberg-atom pairs~\cite{sassmannshausenLongrangeRydbergMolecules2016}. For comparison, photoassociation spectra obtained in the single-species MOTs (gray-dashed lines) are shown. Photoassociation in the dsMOT yields overall larger signals, but only few additional resonances can be identified unambiguously using solely the PFI trace. However, comparison with the signals of molecular ions resulting from autoionization shown in the central and right-hand panels allow us to assign resonances in the PFI spectra to homo- or heteronuclear LRMs. Because of the dense spectra of homo- and heteronuclear LRMs when Cs is excited, many of their photoassociation resonances overlap in the PFI spectrum and the additional information obtained in the autoionization signals is crucial for the assignment.
As an example, the position of the strong resonance in the Cs PFI spectra at \SI{-83}{\mega\hertz} is marked in the upper panels by a green line. This resonance is assigned to \CsK{} based on the observations that (\emph{i}) its intensity is much stronger when photoassociation occurs in the dsMOT and (\emph{ii}) it is also present in the CsK$^+$ signal.

\begin{figure}[tb]
	\centering
	\includegraphics[width=\linewidth]{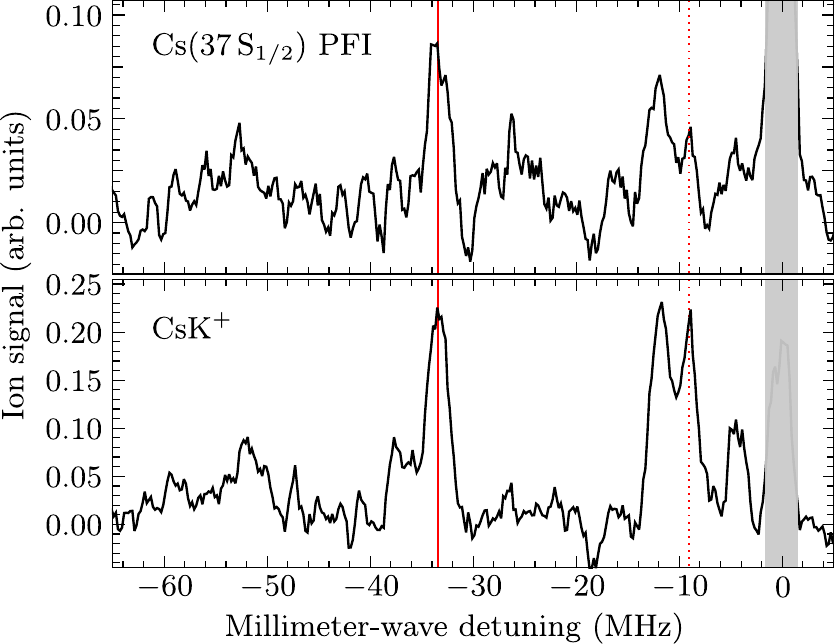}
	\caption{\new{MmW spectrum of the transition from a \lrrmnohf{Cs}{37}{P}{3/2}{K}{4}{S}{1/2}($^{3}\Sigma^+({v=0})$) LRM to \lrrmnohf{Cs}{37}{S}{1/2}{K}{4}{S}{1/2} LRMs (K in $F=2$). (upper panel) Cs(37\,S$_{1/2}$) PFI and (lower panel) CsK$^+$ signal as function of the mmW detuning from the transition \lrrmnohf{Cs}{37}{S}{1/2}{K}{4}{S}{1/2}\,$\leftarrow$ ${^{3}\Sigma^+(v=0)}$.} The resonance assigned to the $v=0$ level in the outermost well of the $^{3}\Sigma^+$ ($^{1,3}\Sigma^+$) state is marked by a full (dotted) red line. The gray bar masks a signal from two-photon excitation of atomic Rydberg states caused by overlapping UV and mmW pulses. The signals have been averaged as in \autoref{fig:spectracomp3129} (resolution \SI{1.0}{MHz}).
		\label{fig:mmW37s}
	}
\end{figure}

MmW spectroscopy of molecules in states with small autoionization rates allows us to probe molecular states correlating to Rydberg states with different angular momenta. The mmW radiation is obtained by frequency-multiplication of the output of a radio-frequency generator (\textsc{Agilent E8257D}), see Appendix~\ref{appendix:mmw}~\cite{peperPrecisionMeasurementIonization2019}. Pulses with length of a few to tens of microseconds and adjustable power are emitted into free space via suitable horns and sent into the vacuum chamber through an optical viewport.
The transitions were detected state-selectively using ramped PFI \cite{peperPrecisionMeasurementIonization2019}. \autoref{fig:mmW37s} shows exemplary mmW spectra of \CsK{} molecules formed by photoassociation of the $^3\Sigma^+(v=0)$ state correlated to the \lrrm{Cs}{37}{P}{3/2}{K}{4}{S}{1/2}{2} asymptote, where transitions to \lrrm{Cs}{37}{S}{1/2}{K}{4}{S}{1/2}{2} molecular states are driven by
\SI{30}{\micro\second} simultaneous photoassociation and mmW pulses. The $n^2$S$_{1/2}$ states of cesium ($\delta_0=4.04$) lie just below the degenerate manifold of high-$l$ Rydberg states and are strongly influenced by LRMs correlated to these asymptotes~\cite{booth2015}. The autoionization signal exhibits several resonances originating from transitions to states with higher autoionization rates. These resonances also appear in the PFI signal for \ars{Cs}{37}{S}{1/2} Rydberg states (upper panel), which confirms mmW-induced population transfer.

To assign observed resonances, we calculate potential energy curves (PECs) of LRMs using the Hamiltonian given by~\citet{eilesHamiltonianInclusionSpin2017a} and the most recent theoretical phase shifts~\cite{eilesFormationLongrangeRydberg2018,khuskivadzeAdiabaticEnergyLevels2002}, see Appendix~\ref{appendix:phaseshifts}. The phase shifts depend on the collision energy, which is taken as the semi-classical kinetic energy of the Rydberg electron
\begin{equation}\label{eq:classicalkinenergy}
E_\mathrm{col}=\frac{2}{R}-\frac{1}{n_\mathrm{ref}^2}
\end{equation}
in the potential of the Rydberg-atom core. The manifold above the state of interest ($n=28$ for the $n\,^2$P states studied here) is chosen as reference $n_\mathrm{ref}$~\cite{eilesHamiltonianInclusionSpin2017a,eilesFormationLongrangeRydberg2018}.

The Hamiltonian is evaluated in an atomic product basis, consisting of the nuclear and electronic spins $(i,m_i,s,m_s)$ of the ground-state atom and a set of states $(n,l,j,m_j)$ of the Rydberg atom, where $m_i$, $m_s$, and $m_j$ are the projections of $i$, $s$, and $j$ on the internuclear axis. The projection of the total angular momentum excluding rotation on the internuclear axis, $\Omega=m_i+m_s+m_j$, is a conserved quantity and we perform the calculation separately for each $\Omega$. The choice of the basis set is ambiguous, because the eigenenergies do not converge when more states are included~\cite{feyComparativeAnalysisBinding2015}. We optimize the basis size separately for each perturber to reproduce the experimental data. The electronic states obtained in our calculations are labeled by the dominant projection of the orbital angular momentum on the internuclear axis, $\Lambda=m_l$, and the spin-character of contributing scattering channels. $^3\Lambda$ states have only binding contributions from scattering channels with a total electronic spin $S=1$, whereas $^{1,3}\Lambda$ states are of mixed $S=0$ and $S=1$ character and have smaller, $F$-dependent binding energies~\cite{Anderson2014}.

\begin{figure}
	\centering
	\includegraphics[width=\linewidth]{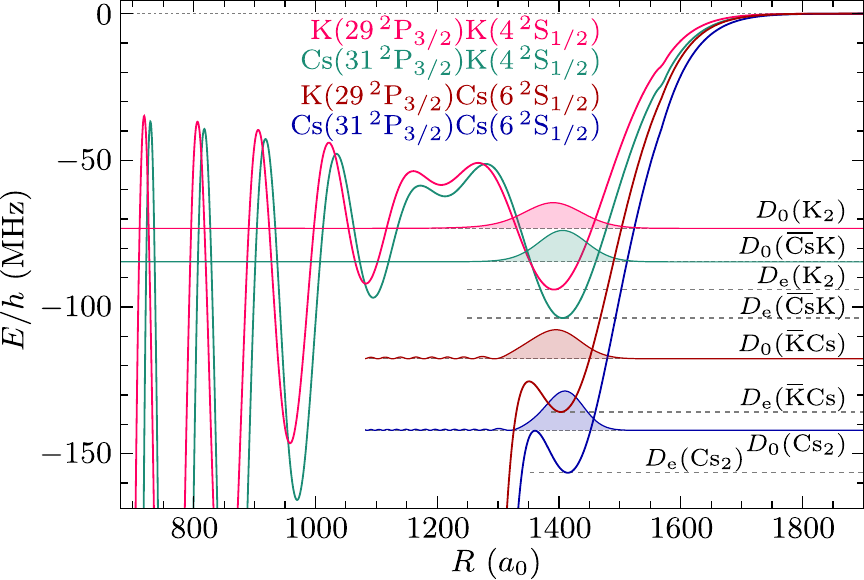}
	\caption{$^3\Sigma^+$($\Omega=\nicefrac{1}{2}$) interaction potentials correlating to the indicated molecular asymptotes. The minima of the outermost potential well $D_\mathrm{e}$ are marked by dashed gray lines. The energy $D_0$ of the vibrational ground states and the respective probability densities are indicated by filled curves.\label{fig:comp31CsKCs2}}
\end{figure}

In \autoref{fig:comp31CsKCs2}, the calculated PECs of $^3\Sigma^+$($\Omega=1/2$) states are drawn for Cs$_2$, K$_2$, \KCs{} and \CsK{}. The PECs of homo- and heteronuclear molecules with the same perturber are similar, because phase shifts and hyperfine interaction of the ground-state atom are the same (see \autoref{eq:isoelectronic}) and the effective quantum numbers of the chosen Cs and K Rydberg states are almost identical. The difference between \CsK{} and K$_2$ (\KCs{} and Cs$_2$) originates mostly from the different atomic spin-orbit splittings.

Molecular states below the \lrrmnohf{Cs}{31}{P}{3/2}{Cs}{6}{S}{1/2} and \lrrmnohf{K}{29}{P}{3/2}{Cs}{6}{S}{1/2} asymptotes are strongly perturbed by the $^3P_0$ shape resonance of the electron-cesium scattering, which causes the steep drop of the PEC at about $R\approx\num{1300}\,a_0$~\cite{khuskivadzeAdiabaticEnergyLevels2002}. The fine-structure splitting of this $^3P_J$ resonance lifts the degeneracy of the PECs for different values of $\Omega$~\cite{deissObservationSpinorbitdependentElectron2020}. However, except for the largest value $\Omega=7/2$, the values for $D_\mathrm{e}$ are within \SI{2}{\mega\hertz} and we report their mean value (see Appendix~\ref{appendix:pecs}). Dissociation energies $D_\mathrm{e}$ of the outermost well of the PECs are summarized in \autoref{tab7:positionscalccs}.

The vibrational $v=0$ levels in the outermost wells of the $^3\Sigma^+$ and $^{1,3}\Sigma^+$ PECs always yield strong photoassociation resonances~\cite{sassmannshausenprl2015} and we focus on these levels. Binding energies and vibrational wavefunctions (see \autoref{fig:comp31CsKCs2}) are determined by the modified Milne phase-amplitude method (see Appendix~\ref{appendix:vibwfcts} and Ref. \cite{sidkyPhaseamplitudeMethodCalculating1999a}) and reported in \autoref{tab7:positionscalccs}. We neglect the rotation of the LRMs, because rotational states are not resolved. The PEC of the $^3\Sigma^+$ state of Cs$_2$ barely supports a vibrational level. In \KCs{}, this level is a resonance bound by quantum scattering at the steep drop of the PEC~\cite{bendkowsky2010}. The calculated width of this resonance is $\Gamma_\mathrm{FWHM}=\SI{7.2}{\mega\hertz}$, in agreement with the observed width of \SI{7.7(7)}{\mega\hertz} (see Appendix~\ref{appendix:vibwfcts}).

\begin{table}[t!]
\renewcommand{\arraystretch}{1.25}
	\caption{Calculated dissociation energies $D_\mathrm{e}$, and calculated ($D_0$) and experimental ($D_\mathrm{0,exp}$) binding energies of the $v=0$ level of the outermost well (in MHz). The ground-state atom is indicated by Cs/K($F$). The experimental uncertainty is estimated from the calibration uncertainty (see text) and the linewidth. The last column contains the deviation of the experimental binding energies relative to the calculated ones.\label{tab7:positionscalccs}}
	\begin{center}
		\begin{tabular}{r c S[table-format=4.1] S[table-format=4.2] S[table-format=1.2] r}
			\hline\hline
		\multicolumn{1}{c}{Asymptote}	&State & \multicolumn{1}{c}{$D_\mathrm{e}/h$} & \multicolumn{1}{c}{$D_0/h$} &  \multicolumn{1}{c}{$D_\mathrm{0,exp}/h$} & \\\hline
		\ars{Cs}{31}{P}{3/2}Cs(3) &$^3\Sigma^+$	& -156.3 &  -141.8 & -134.3(15)&\SI{-5.3}{\percent}\\ \vspace{1mm}
		&$^{1,3}\Sigma^+$&	 -77.3 & -68.8 & -68.6(15)&\SI[retain-explicit-plus]{-0.3}{\percent}\\
		\ars{K}{29}{P}{3/2}Cs(3) &$^3\Sigma^+$	&-135.5 & -117.1 & -114.3(20)&\SI{-2.4}{\percent}\\
		&$^{1,3}\Sigma^+$&	 -71.9 & -59.7 & -59.6(20)&\SI{-0.2}{\percent}\\\hline\hline
			\ars{K}{29}{P}{3/2}K(1) &$^3\Sigma^+$& -94.0 &  -73.2 & -74.5(20)&\SI[retain-explicit-plus]{+1.1}{\percent}\\\vspace{1mm}
			 &$^{1,3}\Sigma^+$& -50.3  & -37.9 & -42.1(20)&\SI[retain-explicit-plus]{+11.3}{\percent}\\
			\ars{Cs}{31}{P}{3/2}K(1) &$^3\Sigma^+$& -103.7 &  -84.6 & -82.7(15)&\SI[retain-explicit-plus]{-2.2}{\percent}\\
	      &$^{1,3}\Sigma^+$& -60.2 & -46.9 & -47.8(15)&\SI[retain-explicit-plus]{+1.9}{\percent}\\\hline
			\ars{Cs}{37}{S}{1/2}K(2) & $^3\Sigma$ & -70.2 & -50.4  &   -33.4(5) &\SI{-33.7}{\percent}\\
		                                           & $^{1,3}\Sigma$ & -26.2 & -15.6&  -9.1(5)  & \SI{-41.7}{\percent}\\\hline\hline
		\end{tabular}
	\end{center}
\end{table}

We assign experimental resonances following three criteria: (\emph{i}) a strong PFI signal, (\emph{ii}) agreement with calculated binding energies, and (\emph{iii}) a comparison between spectra recorded with atoms prepared in different hyperfine states ($^3\Sigma^+$ resonances do not depend on $F$~\cite{Anderson2014,sassmannshausenprl2015}), see Appendix~\ref{appendix:assignment}. The resulting assignment is indicated in \autoref{fig:spectracomp3129}. The \CsK{} resonance at \SI{-83}{\mega\hertz} in the upper panels of Fig.~\ref{fig:spectracomp3129}, for example, is accompanied by a second \CsK{} resonance at a slightly larger detuning of about \SI{-88}{\mega\hertz}. Within the uncertainty of the calculations, both resonances could arise from photoassociation of molecules in the $^{3}\Sigma^+(v=0)$ level. The ratio of signals in the CsK$^+$ and the PFI channel at \SI{-83}{\mega\hertz} is much smaller than at \SI{-88}{\mega\hertz}, indicating a low molecular autoionization rate, as predicted for the $v=0$ level in the outermost well (see Appendix~\ref{appendix:vibwfcts}). We thus assign the resonance at \SI{-83}{\mega\hertz} to $^{3}\Sigma^+(v=0)$.

For the states where Cs is the perturber (Cs$_2$ and \KCs{}), we find the best agreement (\textit{i.e.}, the smallest root mean squared deviation $\sigma_\mathrm{rms}=\SI{4.0}{\mega\hertz}$
) for a basis set including manifolds $n=27 - 28$. Calculation and experiment agree within the experimental uncertainty, except for the $^{3}\Sigma^+$ resonance of Cs$_2$, where the calculation overestimates the binding energy significantly. The experimental assignment of this resonance is unambiguous and the discrepancy increases when increasing the basis set. For $n$P states where K is the perturber, we find the best agreement $\sigma_\mathrm{rms}=\SI{2.4}{\mega\hertz}$ for a basis including manifolds $n=27 - 30$, with a significant deviation for the K$_2$ $^{1,3}\Sigma^+$ resonance. Using the analogous basis set $n=32 - 35$ (one asymptote below, three above the state of interest) for \lrrmnohf{Cs}{37}{S}{1/2}{K}{4}{S}{1/2} overestimates $D_0$ by more than 30~\% (see~\autoref{tab7:positionscalccs}). Even with the minimal basis set ($n=32 - 33$), the calculated binding energies exceed the experimental ones by more than 15~\%. We conclude that a calculation based on the Fermi-pseudo potential and the \textit{ab-initio} calculated \textit{e}-K scattering phase shifts can not yield consistent results for the set of states observed in this work. Only after introducing an \emph{ad-hoc} scaling of the $^3S$ scattering length by 0.951 and choosing the closest-lying atomic Rydberg state as reference  (\autoref{eq:classicalkinenergy}) in a basis including one asymptote below and above the state of interest, we reproduce all experimental binding energies for K$_2$ and \CsK{} within the experimental uncertainty (see Appendix~\ref{appendix:mmw}). When following the same approach for Cs and setting \ars{Cs}{31}{P}{3/2} (\ars{K}{29}{P}{3/2}) as the reference state in the calculations for Cs$_2$ (\KCs{}), we have to reduce the $^3S$ scattering length by about 13~\% to match the experiment. A recent similar modeling of photoassociation resonances in rubidium has yielded a zero-energy $^3S$ scattering length 5~\% smaller than the most recent theoretical calculation~\cite{engelPrecisionSpectroscopyNegativeIon2019,bahrim2001}.

We conclude that (\emph{i}) it seems necessary to set the reference state to the closest-lying atomic state to obtain a consistent description of LRMs correlated to Rydberg states with different angular momentum $l$, (\emph{ii}) for K, Cs, and Rb~\cite{engelPrecisionSpectroscopyNegativeIon2019}, the phase shifts extracted with help of the Fermi-pseudo potential model are significantly smaller than the ones obtained from high-level \textit{ab-initio} calculations, and (\emph{iii}) the extracted phase shifts depend on the basis set chosen for the calculation and should thus be considered effective, model-based parameters and not accurate determinations of the free-electron--atom scattering phase shifts. The set of experimental binding energies for the different homo- and heteronuclear LRMs presented here provides a challenge to the further development of theory, such as a possible extension of the Green's function formalism~\cite{khuskivadzeAdiabaticEnergyLevels2002} or the generalized local frame-transformation theory~\cite{giannakeasGeneralizedLocalFrametransformation2020} to include all relevant spin couplings. Assigning other observed molecular resonances, which could originate from states located at shorter internuclear distances, will yield information about the energy dependence of the scattering phases~\cite{maclennanDeeplyBound24D2019}.

\begin{acknowledgments}
We thank Fr\'ed\'eric Merkt for continued, invaluable support, and Maximilian Beyer for enlightening discussions. JD thanks Matthew Eiles, Chris Greene, and Ilya Fabrikant for inspiring discussions. This work was supported by the ETH Research Grant ETH-22 15-1 and the NCCR QSIT of the Swiss National Science Foundation.
\end{acknowledgments}

\appendix
\section{Density-dependent photoassociation rates of homo- and heteronuclear molecules}
\label{appendix:padensities}

The strengths of homo- and heteronuclear photoassociation resonances depend on the relative ground-state densities of the two species~\cite{eilesFormationLongrangeRydberg2018}. The strength of a photoassociation resonance of $\rm AB$ is proportional to
\begin{equation}\label{eq:hetero_heterorate}
\mathcal{P}_\mathrm{AB}= 4 \uppi V \rho_{\rm A} \rho_\mathrm{B} R_{\mathrm{e},\rm AB}^2  \;,
\end{equation}
and the strength of a resonance of $\rm A_2$ is proportional to
\begin{equation}\label{eq:hetero_homorate}
\mathcal{P}_\mathrm{A_2}= 4 \uppi V \rho_{\rm A}^2 R_{\mathrm{e},\rm A_2}^2  \; ,
\end{equation}
where $V\rho_\mathrm{A}$ is the number of particles of species A in the effective photoassociation volume $V$ and $R_{\mathrm{e}}$ is the equilibrium distance of the photoassociated molecule.

\begin{figure}
	\centering
	\includegraphics[width=\linewidth]{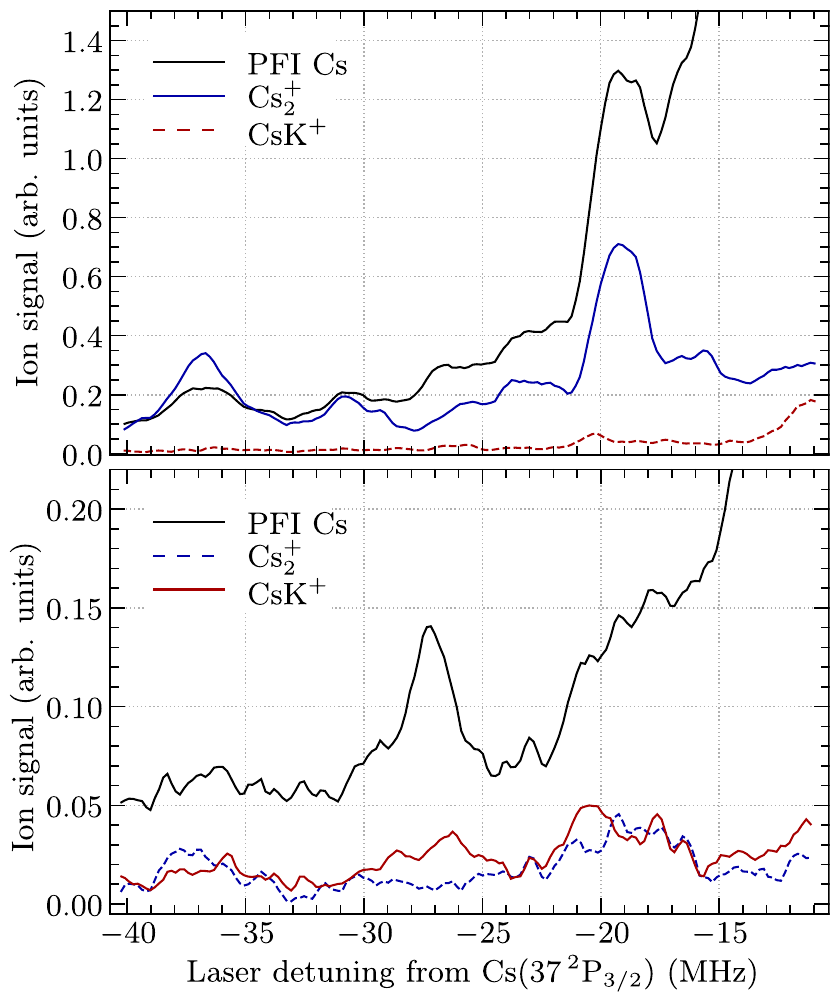}
	\caption{\label{fig:densitydep37}Cs pulsed-field ionization (PFI) signal, Cs$_2^+$, and CsK$^+$ ion signals as a function of the detuning of the photoassociation laser from the \ars{Cs}{37}{P}{3/2}$\leftarrow$\arshf{Cs}{6}{S}{1/2}{3} transition. Prior to photoassociation, the K (Cs) atoms were prepared in the $F=1$ ($F=3$) hyperfine component of the ground state. For the spectra depicted in the upper panel, the Cs and K densities were set to \SI{2.0E10}{\per\centi\meter\cubed} and \SI{1.0E10}{\per\centi\meter\cubed}, respectively. For the spectra depicted in the lower panel, the Cs and K densities were set to \SI{0.4E10}{\per\centi\meter\cubed} and \SI{4.7E10}{\per\centi\meter\cubed}, respectively.}
\end{figure}

The relative strength of the photoassociation signal arising from heteronuclear $\rm \overbar{A}B$ molecules compared to A$_2$ molecules is thus optimized by maximizing the density $\rho_\mathrm{B}$ while keeping the density $\rho_\mathrm{A}$ as small as possible. This density dependence is illustrated in \autoref{fig:densitydep37} for $\overbar{\mathrm{A}}=\overbar{\mathrm{Cs}}$ and $\mathrm{B}=\mathrm{K}$. In the upper panel, the ratio $\rho_{\rm K}$ over $\rho_{\rm Cs}$ is 0.5, whereas in the lower panel the ratio is about 12. The PFI signals recorded for the two cases are remarkably different, with dominant features in the upper panel (dominant Cs density) at detunings of \SI{-37}{\mega\hertz} and \SI{-19}{\mega\hertz}. By comparison with the Cs$_2^+$ signal, these features can indeed be assigned to photoassociation of Cs$_2$ molecules. In contrast, the PFI signal in the lower panel (dominant K density) has a dominant feature at \SI{-27}{\mega\hertz}. This resonance can be assigned to the formation of \CsK{} molecules because a similarly strong feature is absent in the PFI signal when the Cs density is dominant.

The different scaling of homo- and heteronuclear photoassociation rates is clearly visible in Fig.~1. Here, the autoionization signals of CsK$^+$ and Cs$_2^+$ at isolated resonances far detuned from the transition from \arshf{Cs}{6}{S}{1/2}{3} to \ars{Cs}{22}{P}{3/2} are shown for a series of experiments where the potassium density was kept constant at about \SI{2.5(3)E10}{\per\centi\meter\cubed} while the cesium density was varied. As expected from Eqs.~\eqref{eq:hetero_heterorate} and \eqref{eq:hetero_homorate}, the number of heteronuclear ions increases linearly with the cesium density, whereas the number of homonuclear ions depends quadratically on the cesium density. The respective scalings are highlighted by fitting the corresponding linear and quadratic models (Eq.~\eqref{eq:hetero_heterorate} and \eqref{eq:hetero_homorate}) to the data.

\begin{figure*}[tb]
	\centering
	\includegraphics[width=\linewidth]{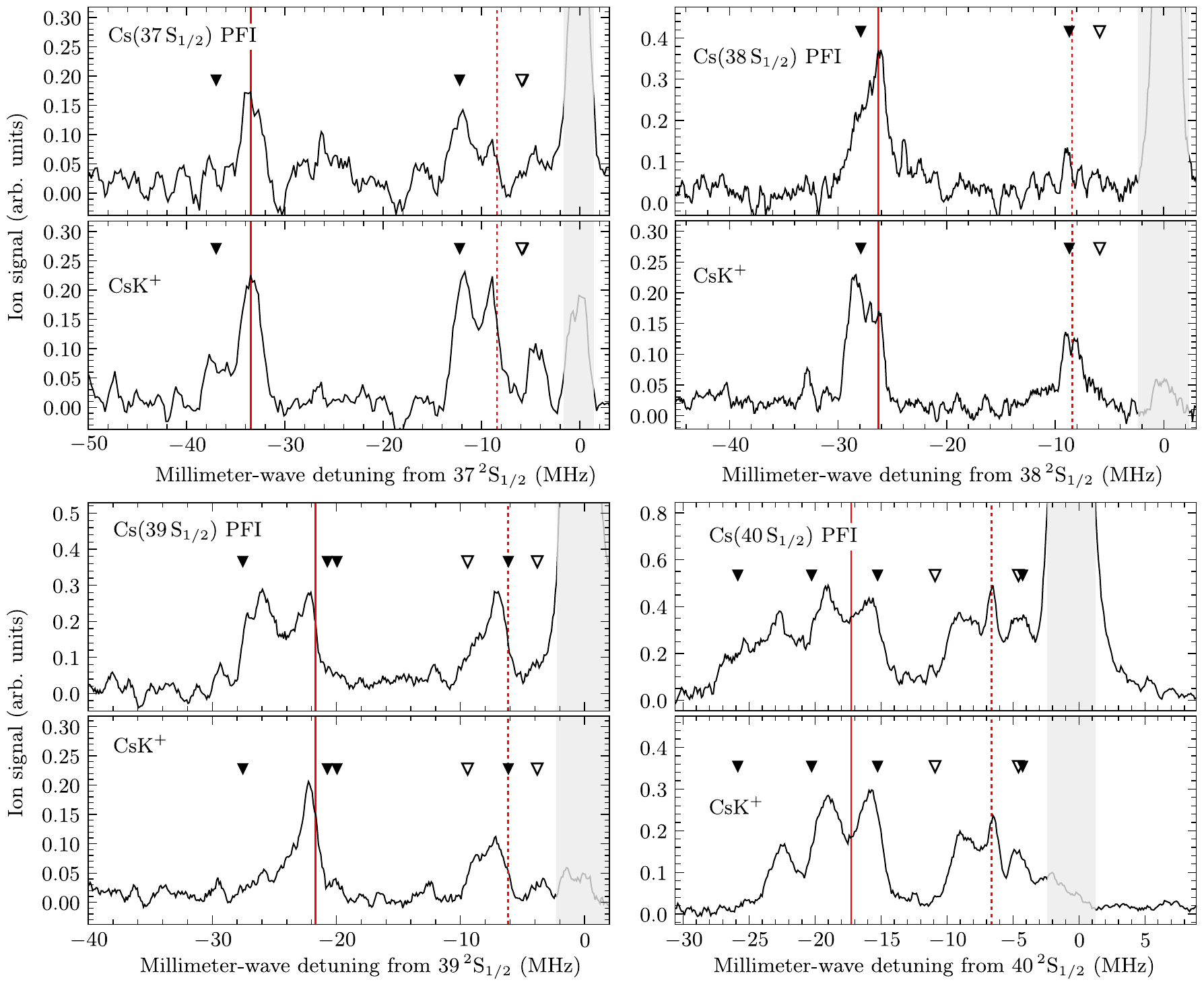}
	\caption{Millimeter-wave spectra of \CsK{} molecules photoassociated in the vibrational ground state ($v=0$) of the outermost potential well of the \lrrmnohf{Cs}{37}{P}{3/2}{K}{4}{S}{1/2} $^3\Sigma^+$ interaction potential. The initial state was photoassociated by a \SI{30}{\micro\second} UV laser pulse from a sample of Cs (K) atoms prepared in $F=4$ ($F=2$) hyperfine state. Using state-selective pulsed field ionization, the $n\,\mathrm{S}_{1/2}$ and CsK$^+$ ion signal was recorded as a function of the millimeter-wave detuning from the (upper left) \lrrm{Cs}{37}{S}{1/2}{K}{4}{S}{1/2}{2}, (upper right) \lrrm{Cs}{38}{S}{1/2}{K}{4}{S}{1/2}{2}, (bottom left) \lrrm{Cs}{39}{S}{1/2}{K}{4}{S}{1/2}{2}, and (bottom right) \lrrm{Cs}{40}{S}{1/2}{K}{4}{S}{1/2}{2} asymptotes. The calculated positions (see text for details) of the vibrational ground state in the outermost potential wells of the $^3\Sigma^+$ ($^{1,3}\Sigma^+$) electronic state are indicated by full (dashed) red lines. The position of further bound states with overlapping initial and final vibrational wavefunctions are indicated by full (empty) triangles. Note that the calculations have been performed with \emph{ad-hoc} scaled parameters.
		\label{fig:mmW3740s}
	}
\end{figure*}

\section{Millimeter-wave spectroscopy and assignment of resonances}
\label{appendix:mmw}

\begin{figure}
	\centering
	\includegraphics[width=\linewidth]{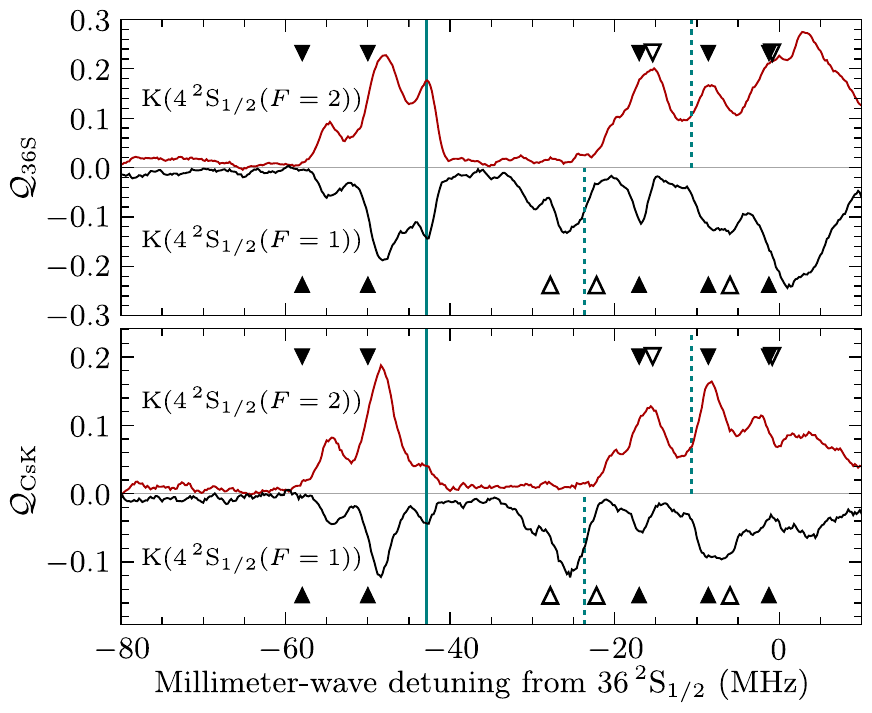}
	\caption{Millimeter-wave spectra of \CsK{} molecules photoassociated in the vibrational ground state of the outermost potential well of the \lrrmnohf{Cs}{34}{P}{3/2}{K}{4}{S}{1/2} $^3\Sigma^+$ state by a \SI{30}{\micro\second} UV laser pulse. Before photoassociation, the \potassium{} atoms were prepared either in the $F=1$ (black) or $F=2$ (red) hyperfine component of the ground-state atom. The upper panel depicts the ratio $\mathcal{Q}_{36\mathrm{S}}$ of transferred Rydberg atoms and the lower panel depicts the ratio $\mathcal{Q}_\mathrm{CsK}$ after a millimeter-wave pulse of \SI{10}{\micro\second} length with a frequency close to the \ars{Cs}{36}{S}{1/2}$\leftarrow$\ars{Cs}{34}{P}{3/2} transition.
	For clarity, the sign of the black trace is inverted and constant offsets of 0.12 and 0.13 (arb. units) were subtracted from $\mathcal{Q}_\mathrm{36S}$ and $\mathcal{Q}_\mathrm{CsK}$, respectively. The calculated positions (see text for details) of the $v=0$ level in the outermost potential well of the $^3\Sigma^+$ ($^{1,3}\Sigma^+$) electronic state are indicated by full (dashed) cyan lines. The position of further vibrationally bound states with overlapping vibrational wavefunctions of the initial and final molecular state are indicated by full and empty triangles for $^3\Sigma^+$ and $^{1,3}\Sigma^+$, respectively. Note that the calculations have been performed with \emph{ad-hoc} scaled parameters.
	\label{fig:mmW36s}}
\end{figure}

The lifetimes of long-range Rydberg molecules can exceed several microseconds, making it possible to perform millimeter-wave spectroscopy of these molecules. Millimeter-wave transitions allow one to probe molecular states which are inaccessible in direct one-photon photoassociation such as molecular states lying below $n\,^2\mathrm{S}_{1/2}$ asymptotes. To this end, molecules are formed by photoassociation with UV-laser pulses of typically \SI{30}{\micro\second} length. Millimeter-wave radiation is applied to the sample simultaneously with  (\autoref{fig:mmW3740s}) or after (\autoref{fig:mmW36s}) the UV radiation.

The mmW radiation is obtained by frequency-multiplication of the output of a radio-frequency generator (\textsc{Agilent E8257D})~\cite{peperPrecisionMeasurementIonization2019} using active 6-fold  (70--\SI{110}{GHz}, \textsc{OML Inc. S10MS-AG}), 18-fold (170--\SI{250}{GHz}, \textsc{Virginia Diodes WR-9.0} with \textsc{Virginia Diodes WR4.3x2}) or 27-fold (220--\SI{330}{GHz}, \textsc{Virginia Diodes WR-9.0} with \textsc{Virginia Diodes WR3.4x3}) multipliers. Pulses with length of a few to tens of microseconds and adjustable power were formed by modulating the power of the fundamental either using an internal pulse generator of the rf source or using an absorptive modulator (\textsc{Hewlett Packard 33001C}) driven by an arbitrary waveform generator. After attenuation through an adjustable, calibrated attenuator, the millimeter-wave radiation is emitted into free space via suitable horns and sent into the vacuum chamber through an optical viewport. Additional sheets of paper are used to further attenuate the millimeter-wave radioation. The power of the millimeter-wave radiation at the position of the atom cloud is not well known. However, over the spectral region of the millimeter-wave spectra presented here, intensity modulations resulting from reflections off the metallic walls of the vacuum chamber can be neglected.

Subsequently, a ramped field-ionization pulse is applied and the Rydberg states are detected state-selectively. The state-selective detection of Rydberg molecules relies on the observation that the electronic wavefunction of low-$l$ Rydberg molecules is dominated by the atomic character of the molecular dissociation asymptote. The molecules thus ionize at an electric field closely corresponding to the field at which the corresponding Rydberg atom would ionize. This property allows us to detect the transition from the vibrational ground state $v=0$ in the $^3\Sigma^+$ state associated with the \lrrmnohf{Cs}{37}{P}{3/2}{K}{4}{S}{1/2} asymptote to molecules in states associated with the \lrrmnohf{Cs}{n}{S}{1/2}{K}{4}{S}{1/2} asymptotes. Millimeter-wave spectra of \lrrm{Cs}{n}{S}{1/2}{K}{4}{S}{1/2}{2}$\leftarrow$\lrrm{Cs}{37}{P}{3/2}{K}{4}{S}{1/2}{2} transitions with $n=37,...,40$ are depicted in \autoref{fig:mmW3740s}. The upper panel depicts the Cs($n\,^2\mathrm{S}_{1/2}$) PFI and the lower panel depicts the CsK$^+$ ion signal as a function of the millimeter-wave frequency detuning from the respective atomic asymptote. The spectra show distinct peaks, both in the Cs($n\,^2\mathrm{S}_{1/2}$) PFI and the CsK$^+$ signals, indicating bound-bound millimeter-wave transitions. We assign these features on the basis of the Franck-Condon principle, which allows only for transitions between states with overlapping vibrational wavefunctions of the final and initial molecular state.

\begin{figure}
	\centering
	\includegraphics[width=\linewidth]{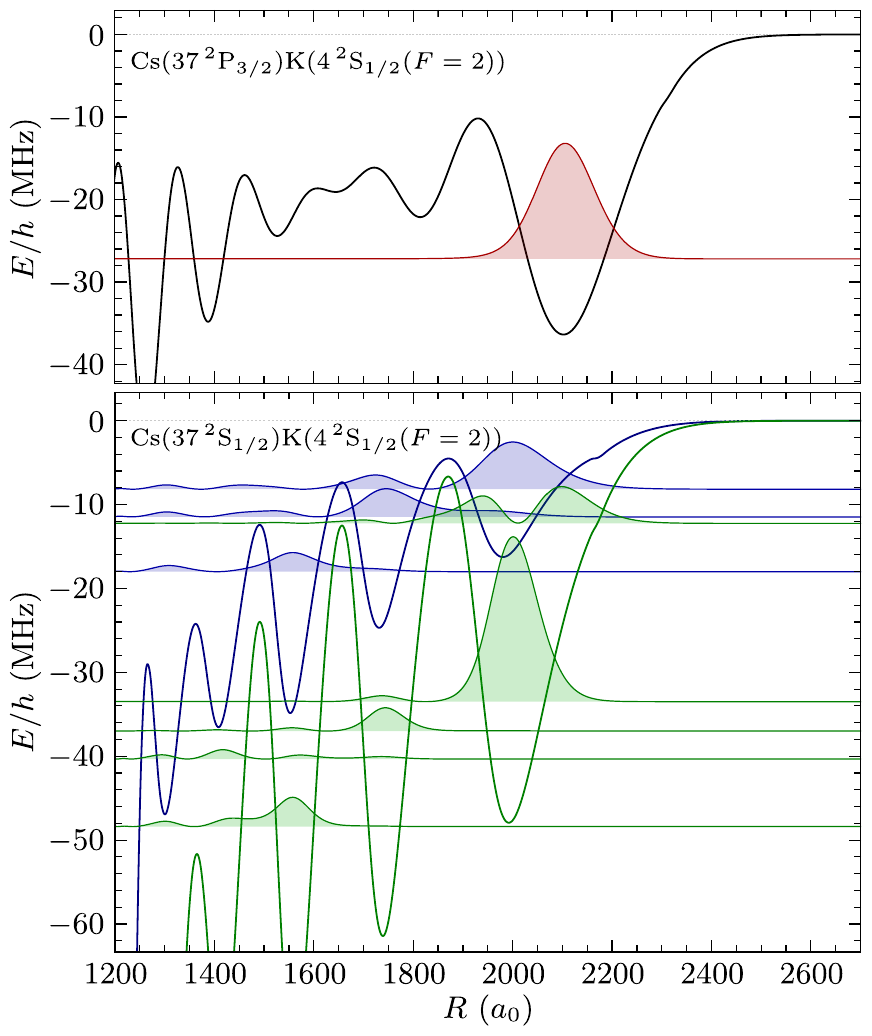}
	\caption{(Upper panel) $^3\Sigma^+$ PEC (black)  correlating to the \lrrm{Cs}{37}{P}{3/2}{K}{4}{S}{1/2}{2} asymptote. The energy and probability density of the vibrational ground state $v=0$ in the outermost potential is given by a red filled line. (Lower panel) Calculated PECs for $^3\Sigma^+$ (green) and $^{1,3}\Sigma^+$ (blue) states correlating to the \lrrm{Cs}{37}{S}{1/2}{K}{4}{S}{1/2}{2} asymptote. The energy and probability density of the vibrational levels in the respective potentials is given by green and blue filled lines. Note that the calculations have been performed with \emph{ad-hoc} scaled parameters.
		\label{fig:potential37s}
	}
\end{figure}

For example, the molecular interaction potentials of the states involved in the \lrrm{Cs}{37}{S}{1/2}{K}{4}{S}{1/2}{2}$\leftarrow$\lrrm{Cs}{37}{P}{3/2}{K}{4}{S}{1/2}{2} transition are depicted in \autoref{fig:potential37s}. The \lrrm{Cs}{37}{P}{3/2}{K}{4}{S}{1/2}{2} $^3\Sigma^+$ interaction potential of the initial level is shown in the upper panel in black and the position and probability density of the vibrational ground state in the outermost potential well is given by a red filled line. The wavefunction of the initial state only has significant overlap with the outermost potential well of the \lrrm{Cs}{37}{S}{1/2}{K}{4}{S}{1/2}{2} $^3\Sigma^+$ (green) and $^{1,3}\Sigma^+$ (blue) interaction potential and therefore we assign the feature  at \SI{-33.4}{MHz} in the \lrrm{Cs}{37}{S}{1/2}{K}{4}{S}{1/2}{2}$\leftarrow$\lrrm{Cs}{37}{P}{3/2}{K}{4}{S}{1/2}{2} millimeter-wave spectrum to a transition to the vibrational ground state of the \lrrm{Cs}{37}{S}{1/2}{K}{4}{S}{1/2}{2} $^3\Sigma^+$ interaction potential.

The millimeter-wave spectra become more congested with increasing $n$, because the overlap of the initial \lrrm{Cs}{37}{P}{3/2}{K}{4}{S}{1/2}{2} $^3\Sigma^+(v=0)$ bound state with other potential wells increases. A similar effect can be observed for the \lrrmnohf{Cs}{36}{S}{1/2}{K}{4}{S}{1/2}$\leftarrow$\lrrmnohf{Cs}{34}{P}{3/2}{K}{4}{S}{1/2} millimeter-wave transition. The spectra recorded with the potassium ground-state atom prepared in the $F=1$ and $F=2$ hyperfine state are presented in \autoref{fig:mmW36s}. In order to remove noise caused by fluctuations in the number of photoassociated molecules from millimeter-wave spectra, we consider the ratio $\mathcal{Q}_\mathrm{36S}$($\mathcal{Q}_\mathrm{CsK}$) of the signal detected in the integration window of the $36\,^2\mathrm{S}_{1/2}$ PFI (CsK$^+$) signal over the PFI ion signal of the initial state. The spectra reveal several transitions to vibrational bound states including a triplet between \SI{-55}{MHz} and \SI{-40}{MHz}, which by comparison of the spectra recorded with the potassium ground-state atoms prepared in the $F=1$ and $F=2$ state can be assigned to states bound in the $^3\Sigma^+$ state. By comparison with calculations, this triplet can be assigned to transitions to the vibrational ground states $v=0$ of the three outermost potential wells correlated to \lrrmnohf{Cs}{36}{S}{1/2}{K}{4}{S}{1/2} (compare also \autoref{fig:potential37s}).

As mentioned in the main article, the basis set and scattering phase shifts used to reproduce the \lrrmnohf{K}{29}{P}{3/2}{K}{4}{S}{1/2} and \lrrmnohf{Cs}{31}{P}{3/2}{K}{4}{S}{1/2} binding energies systematically overestimate the binding energies of the vibrational ground states bound below the \lrrmnohf{Cs}{n}{S}{1/2}{K}{4}{S}{1/2}{2} asymptotes. Therefore, we calculate the interaction potentials in a basis comprising the manifolds $n=(n_\mathrm{S}-5)...(n_\mathrm{S}-4)$ and scale the e-K $^3\mathrm{S}$ scattering length by 0.951 to reproduce the experimental binding energy of the vibrational ground state of the \lrrm{Cs}{37}{S}{1/2}{K}{4}{S}{1/2}{2} $^3\Sigma^+$ interaction potential.

\begin{table}[t]
\renewcommand{\arraystretch}{1.25}
	\caption{Calculated ($D_0$) and experimental ($D_\mathrm{0,exp}$) binding energies of the $v=0$ level of the outermost well using the \emph{ad-hoc} scaled $^3S$ scattering length. The ground-state atom is indicated by Cs/K($F$). The experimental uncertainty is estimated from the calibration uncertainty (see text) and the line width. \label{tab7:potassiumscaled}}
	\begin{center}
		\begin{tabular}{r c S[table-format=5.3] S[table-format=4.4]}
			\hline\hline
		\multicolumn{1}{c}{\multirow{2}{*}{Asymptote}}	&\multirow{2}{*}{State} & \multicolumn{1}{c}{$D_\mathrm{0}/h$} & \multicolumn{1}{c}{$D_0,\mathrm{exp}/h$ }\\
		\multicolumn{2}{c}{}&\multicolumn{2}{c}{(MHz)}  \\\hline
		\lrrmnohf{Cs}{36}{S}{1/2}{K}{4}{S}{1/2} &$^3\Sigma^+$	& -42.9 &  -43.0(5) \\
		&$^{1,3}\Sigma^+(F=1)$&	 -23.7 & -25.7(5) \\\vspace{1mm}
		&$^{1,3}\Sigma^+(F=2)$&	 -10.5 & \textemdash \\
		\lrrmnohf{Cs}{37}{S}{1/2}{K}{4}{S}{1/2} &$^3\Sigma^+$	& -33.5 &  -33.4(5) \\ \vspace{1mm}
		&$^{1,3}\Sigma^+(F=2)$&	 -8.2 & -9.1(5) \\
		\lrrmnohf{Cs}{38}{S}{1/2}{K}{4}{S}{1/2} &$^3\Sigma^+$	& -26.3 &  -26.4(5) \\ \vspace{1mm}
		&$^{1,3}\Sigma^+(F=2)$&	 -8.4 & -8.4(5) \\
		\lrrmnohf{Cs}{39}{S}{1/2}{K}{4}{S}{1/2} &$^3\Sigma^+$	& -21.7 &  -22.2(5) \\ \vspace{1mm}
		&$^{1,3}\Sigma^+(F=2)$&	 -6.2 & -7.2(5) \\
		\lrrmnohf{Cs}{40}{S}{1/2}{K}{4}{S}{1/2} &$^3\Sigma^+$	& -17.3  &  -19.1(5) \\ \vspace{1mm}
		&$^{1,3}\Sigma^+(F=2)$&	 -6.6 & \textemdash \\\hline\hline
		\lrrmnohf{Cs}{31}{P}{3/2}{K}{4}{S}{1/2} &$^3\Sigma^+$	& -83.9 &  -82.7(15) \\ \vspace{1mm}
		&$^{1,3}\Sigma^+(F=1)$&	 -47.5 & -47.8(15) \\
		\lrrmnohf{K}{29}{P}{3/2}{K}{4}{S}{1/2} &$^3\Sigma^+$	& -74.9 &  -74.5(20) \\ \vspace{1mm}
		&$^{1,3}\Sigma^+(F=1)$&	 -40.9 & -42.1(20) \\\hline\hline
		\end{tabular}
	\end{center}
\end{table}

\begin{figure}
	\centering
	\includegraphics[width=\linewidth]{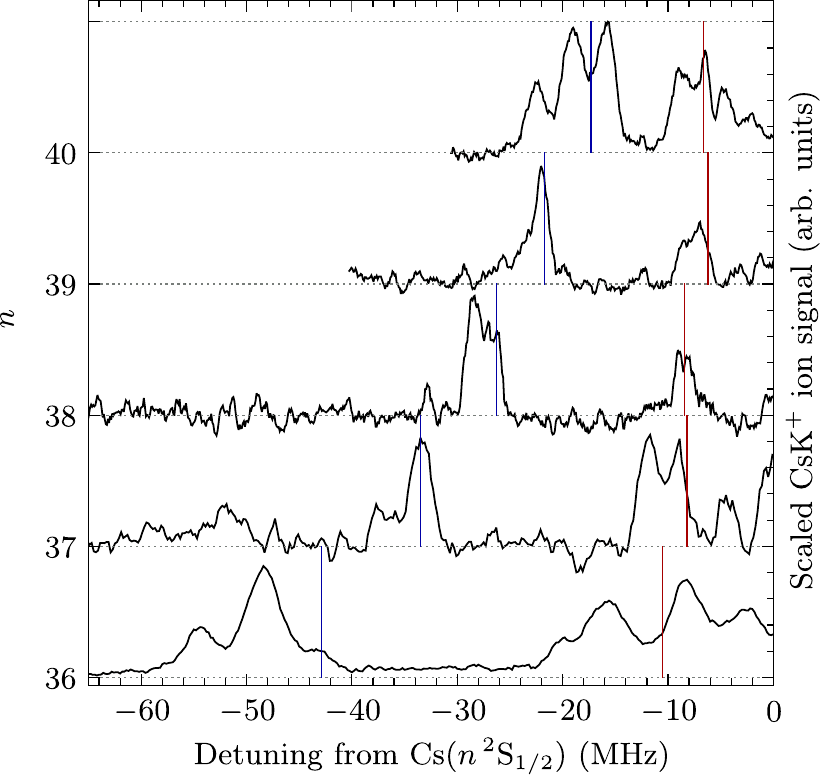}
	\caption{Comparison of CsK$^+$ ion signals recorded as a function of the millimeter-wave radiation from the \lrrm{Cs}{n}{S}{1/2}{K}{4}{S}{1/2}{2} asymptote for $n=36$--40 from bottom to top respectively. For clarity the traces are shifted corresponding to their principal quantum number $n$.
		\label{fig:compmmW}
	}
\end{figure}

The resulting binding energies of the vibrational ground-states in the outermost potential well of the $^3\Sigma^+$ ($^{1,3}\Sigma^+$) interaction potential are depicted by red full (dashed) lines in \autoref{fig:mmW3740s} and summarized in \autoref{tab7:potassiumscaled}. The chosen scaling of the $^3\mathrm{S}$ scattering length reproduces the position of features assigned to the $v=0$ level of the $^3\Sigma^+$ and the $^{1,3}\Sigma^+$ state for \CsK{} molecules associated to \lrrmnohf{Cs}{n}{S}{1/2}{K}{4}{S}{1/2}, \lrrmnohf{Cs}{31}{P}{3/2}{K}{4}{S}{1/2}, and K$_2$ molecules associated to the \lrrmnohf{K}{29}{P}{3/2}{K}{4}{S}{1/2} asymptote. Significant deviations are only observed for \lrrmnohf{Cs}{n}{S}{1/2}{K}{4}{S}{1/2} with $n\geq39$.

To highlight the $n$-scaling of binding energies, \autoref{fig:compmmW} shows the autoionization signals from \autoref{fig:mmW3740s} and \autoref{fig:mmW36s} in a single graph together with the calculated binding energies of the \lrrm{Cs}{n}{S}{1/2}{K}{4}{S}{1/2}{2}\,$^3\Sigma^+(v=0)$ states with $36\leq n\leq40$. The calculated binding energies are found to scale with $1/n^{*7.7}$, slightly deviating from the expected $1/n^{*6}$ scaling~\cite{sassmannshausenprl2015}. We attribute the deviation to the mixing with the close-lying hydrogenic manifold and the contribution from zero-point energies. The binding energies for the $^{1,3}\Sigma^+(v=0)$ state do not scaling systematically with $n$, because the vibrational wavefunctions of these states are not localized purely in the outer-most well (see \textit{e.g.} \autoref{fig:potential37s}).

\section{Scattering phase shifts}
\label{appendix:phaseshifts}

For the calculation of $\mathrm{e}^-$--Cs scattering phase shifts, we follow the procedure described by Khuskivadze \textit{et al.}~\cite{khuskivadzeAdiabaticEnergyLevels2002} using a model potential $V_{LS}(r)$ dependent on the angular momentum $L$ and spin $S$ of the $\mathrm{e}^-$--Cs system. $r$ is the electron--alkali-metal-atom distance.

\begin{figure}
	\centering
	\includegraphics[width=\linewidth]{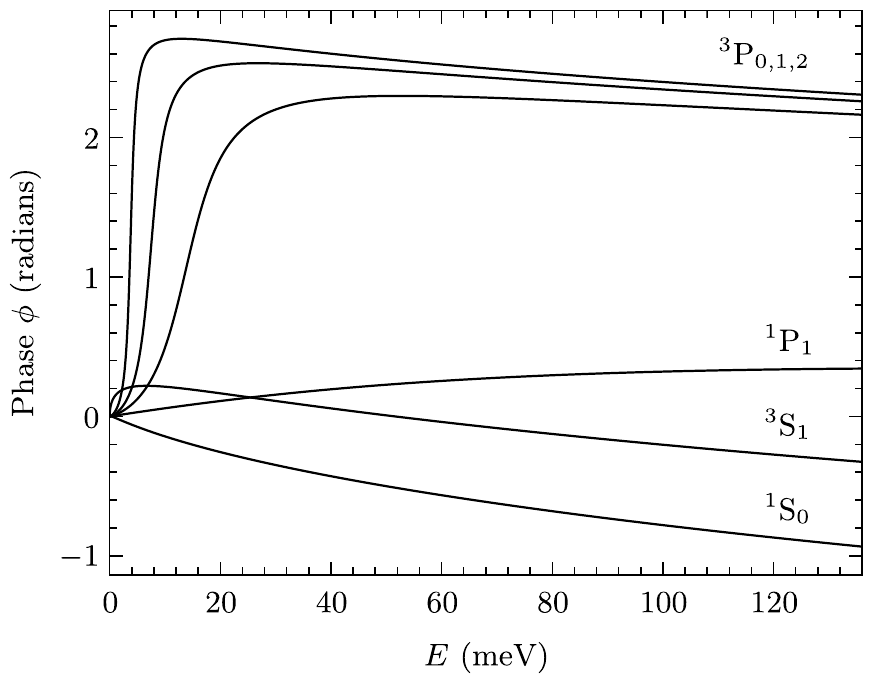}
	\caption{Phase shifts of the e$^-$--Cs scattering process as a function of the electron kinetic energy $E$. The phase shifts for the electron collision with Cs are obtained by the method described in Ref.~\cite{khuskivadzeAdiabaticEnergyLevels2002} including spin-orbit coupling. \label{fig:phaseshifts}
	}
\end{figure}

The resulting coupled set of differential equations is solved numerically using the renormalized Numerov method \cite{johnsonRenormalizedNumerovMethod1978} with a step size of \linebreak$h=\num{5e-4}\,a_0$. At the end points of the integration, $r_\mathrm{A}$ and $r_\mathrm{end}=r_\mathrm{A}+h=2000\,a_0$, the wavefunction is matched to Riccati-Bessel functions \cite{beyerObservationCalculationQuasibound2016}. The phase shifts obtained for the case of the e$^-$--Cs scattering with collision energies between \SI{0}{\milli\electronvolt} and \SI{140}{\milli\electronvolt} are depicted in \autoref{fig:phaseshifts}. Our results agree with the results reported in Ref.~\cite{khuskivadzeAdiabaticEnergyLevels2002} within the resolution of their Figure 3.

The scattering phase shifts for the $\mathrm{e}^-$--K collision have been calculated and reported by Eiles in Ref.~\cite{eilesFormationLongrangeRydberg2018}, Figure 1, from where we extract the phase shifts graphically. In his calculation, spin-orbit effects are neglected. From the $^3\mathrm{P}_0$--$^3\mathrm{P}_1$ fine-structure splitting in Cs$^-$ (\SI{3.78}{meV}) and its scaling with the nuclear charge $Z$ and the principal quantum number $n$ (4 for K and 6 for Cs) as $\nicefrac{Z^4}{n^3}$ \cite{bahrimNegativeionResonancesCross2001}, the $^3\mathrm{P}_0$--$^3\mathrm{P}_1$ fine-structure splitting in K$^-$ is estimated to be \SI{182}{\micro\electronvolt}. Neglecting the spin-orbit interaction in $\mathrm{e}^-$--K scattering is therefore reasonable and hence we treat all three $^3\mathrm{P}_{0,1,2}$ scattering phase shifts to be equal. Because the $^3\mathrm{P}_J$ shape resonance lies at much higher collision energies in e$^-$--K than in e$^-$--Cs collisions (\SI{24}{\milli\electronvolt} vs \SI{8}{\milli\electronvolt} \cite{eilesFormationLongrangeRydberg2018}), the resulting perturbations occur at shorter internuclear distances in K$_2$ than in Cs$_2$.

\section{Calculation of interaction potentials}
\label{appendix:pecs}

We calculate the potential-energy curves (PEC) for homo- and heteronuclear long-range Rydberg molecules using the Hamiltonian given by~\citet{eilesHamiltonianInclusionSpin2017a}. This Hamiltonian explicitly includes the Fermi-contact interaction in a basis of good quantum numbers for the scattering problem $\beta$ (scattering partial wave $L$, total electronic spin $S$, total angular momentum excluding nuclear spin $J$, and projection of $J$ on the internuclear axis) in the pseudo-potential form
\begin{equation}
    \hat{V}_\mathrm{FC}=\sum_\beta\ket{\beta}\frac{\left(2 L +1\right)^2}{2}a\left(SLJ,k\right)\frac{\delta(X)}{r^{2(L+1)}}\bra{\beta},
\end{equation}
where the $a\left(SLJ,k\right)$ scattering parameters depend on the semi-classical wavevector $k$ of the Rydberg electron at the position of the ground-state atom (see Eq. \eqref{eq:classicalkinenergy}), and $r$ is the distance between Rydberg and ground-state atom. Via a frame-transformation matrix, the operator $\hat{V}_\mathrm{FC}$ is expressed in a basis of good quantum numbers $\left(n l j m_j\right)$ of the Rydberg atom, in which the operator $\hat{H}_{0+\mathrm{SO}}$ for the interaction of the Rydberg electron with the ion core is diagonal with eigenvalues given by experimentally determined quantum defects~\cite{deiglmayrpra2016,goypra1982,peperPrecisionMeasurementIonization2019}. The hyperfine interaction of the ground-state atom is included by the effective Hamiltonian
\begin{equation}
    \hat{H}_\mathrm{HF}=A_\mathrm{HF}\, \vec{i}\cdot\vec{s},
\end{equation}
where $A_\mathrm{HF}$ is the hyperfine coupling constant and $\vec{i}$ and $\vec{s}$ represent the nuclear and electronic spin of the ground-state atom, respectively. We neglect the hyperfine interaction in the Rydberg atom, which scales as $n^{-3}$, and the interaction of the positively charged Rydberg core with the ground-state atom, because both contributions are typically too small to be observed at the resolution ($\sim$\SI{1.5}{MHz}) of our experiments.

The total Hamiltonian $\hat{H}= \hat{H}_{0+\mathrm{SO}}+\hat{H}_\mathrm{HF}+\hat{V}_\mathrm{FC}$ is brought into matrix form in an atomic product basis, as described in the main article. It was shown that the pseudo-potential approach~\cite{eilesHamiltonianInclusionSpin2017a} with a basis set including two degenerate manifolds above and below the asymptote of interest reproduces the results of basis-set independent Greens function calculations~\cite{feyComparativeAnalysisBinding2015,engelPrecisionSpectroscopyNegativeIon2019}. With this choice of basis, our calculations underestimate binding energies when K is the perturber and overestimate binding energies when Cs is the perturber and we thus treat the basis size as adjustable parameter of the calculation.
Numerical diagonalization for different internuclear distances $R$ yields the PECs. We have verified that our calculation reproduces the results (where potassium is the perturber) of Figures 3, 4, and 7 of Ref.~\cite{eilesFormationLongrangeRydberg2018} within the line width of the respective curves.

\begin{figure}
	\includegraphics[width=\linewidth]{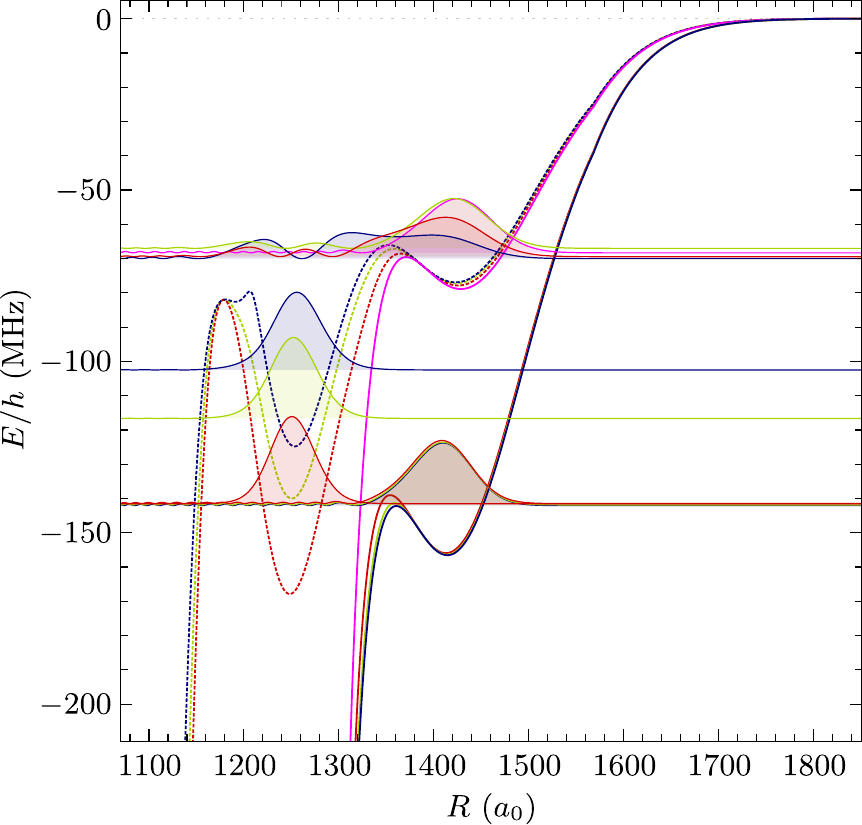}
	\caption{$^3\Sigma^+$($\Omega$) (full) and $^{1,3}\Sigma^+$($\Omega$) (dashed) interaction potentials correlating to the \lrrm{Cs}{31}{P}{3/2}{Cs}{6}{S}{1/2}{3} asymptote with $\Omega=1/2$ (blue), $\Omega=3/2$ (green), $\Omega=5/2$ (red), and $\Omega=7/2$ (pink). The vibrational ground-states ($v=0$) in the outermost potential wells are given by full, filled lines.
		\label{fig:allOmegaCs2}
	}
\end{figure}

\begin{figure}
	\centering
	\includegraphics[width=\linewidth]{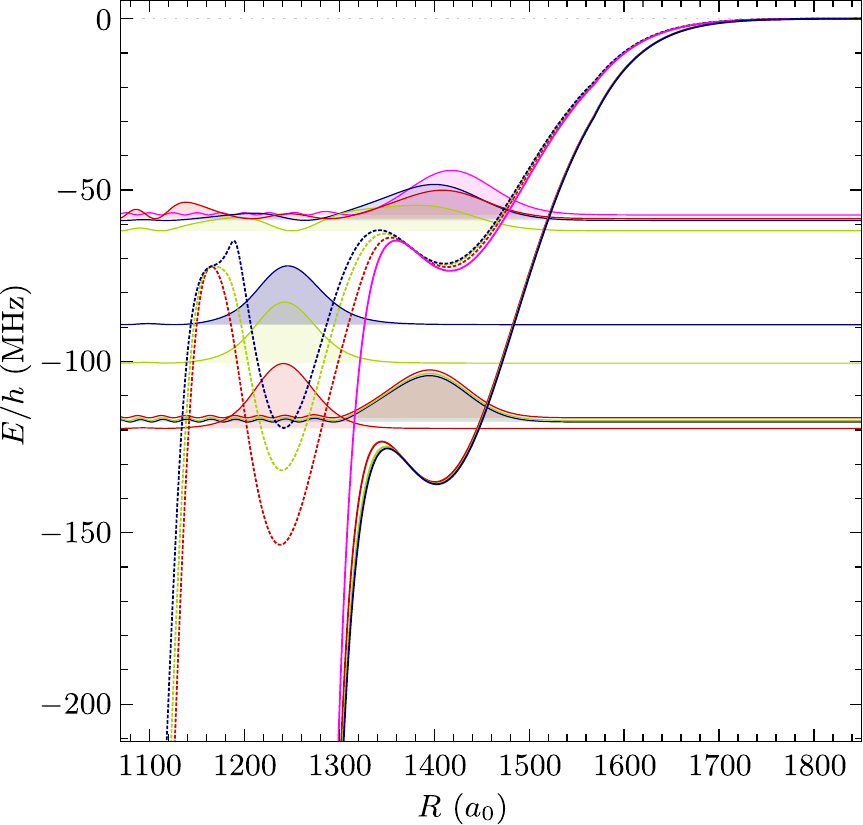}
	\caption{$^3\Sigma^+$($\Omega$) (full) and $^{1,3}\Sigma^+$($\Omega$) (dashed) interaction potentials correlating to the \lrrm{K}{29}{P}{3/2}{Cs}{6}{S}{1/2}{3} asymptote with $\Omega=1/2$ (blue), $\Omega=3/2$ (green), $\Omega=5/2$ (red), and $\Omega=7/2$ (pink). The vibrational ground-states ($v=0$) in the outermost potential wells are given by full, filled lines.
		\label{fig:allOmegaKCs}
	}
\end{figure}

\begin{figure}
	\centering
	\includegraphics[width=\linewidth]{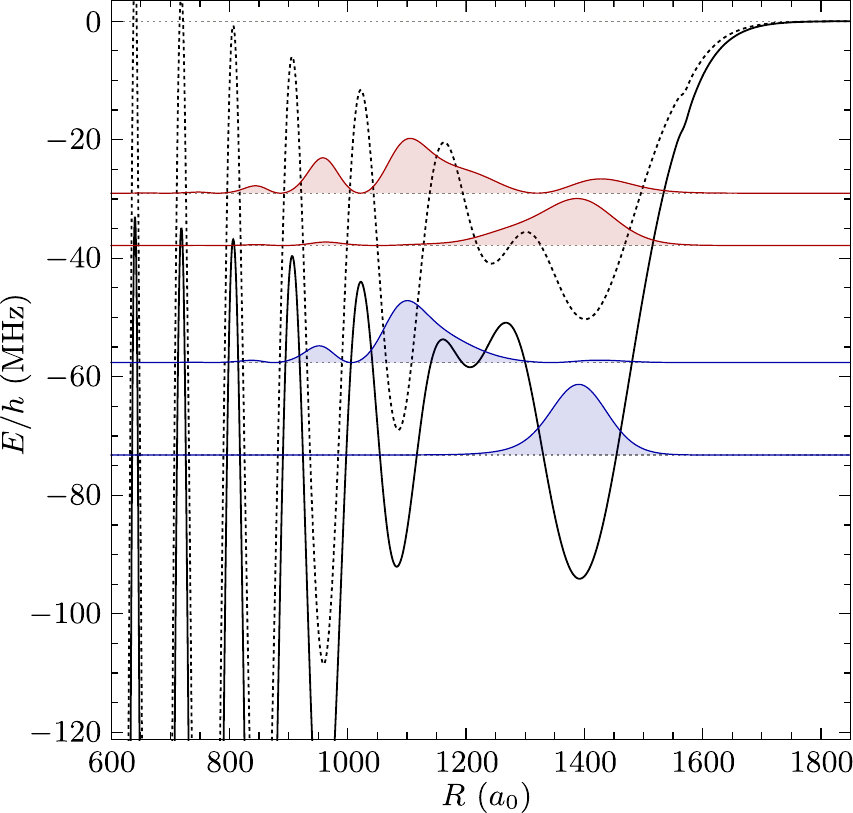}
	\caption{$^3\Sigma^+$ (full black) and $^{1,3}\Sigma^+$ (dashed black) interaction potentials correlating to the \lrrm{K}{29}{P}{3/2}{K}{4}{S}{1/2}{1} asymptote. The vibrational ground-states ($v=0$) in the two outermost potential wells are given by blue ($^3\Sigma^+$) and red ($^{1,3}\Sigma^+$) full, filled lines.
		\label{fig:allOmegaK2}
	}
\end{figure}

\begin{figure}
	\centering
	\includegraphics[width=\linewidth]{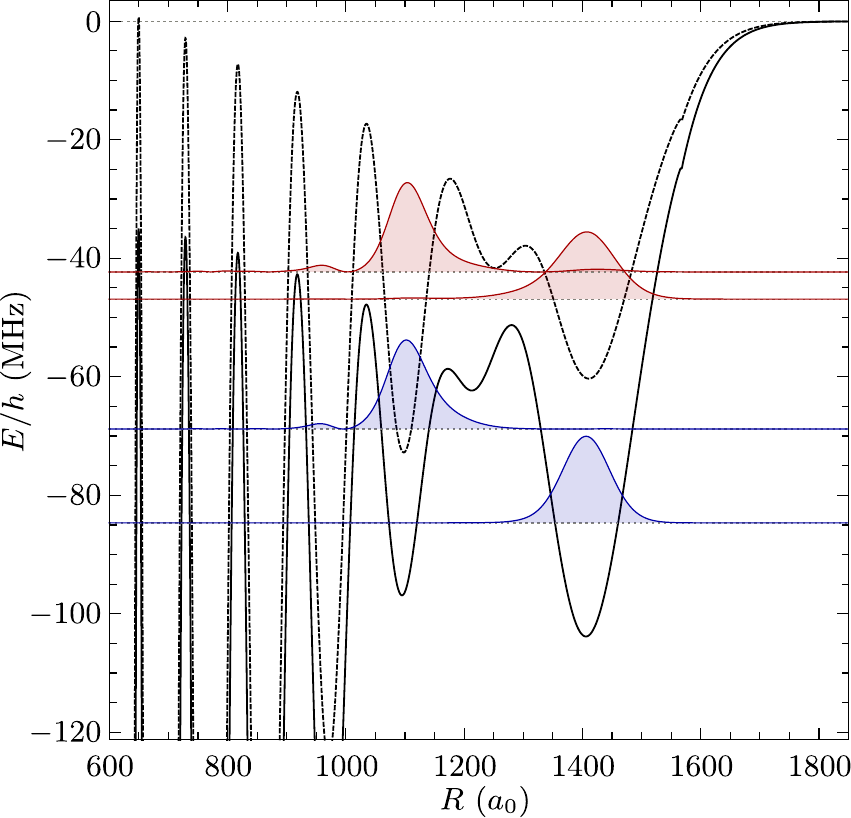}
	\caption{$^3\Sigma^+$ (full black) and $^{1,3}\Sigma^+$ (dashed black) interaction potentials correlating to the \lrrm{Cs}{31}{P}{3/2}{K}{4}{S}{1/2}{1} asymptote. The vibrational ground-states ($v=0$) in the two outermost potential wells are given by blue ($^3\Sigma^+$) and red ($^{1,3}\Sigma^+$) full, filled lines.
		\label{fig:allOmegaCsK}
	}
\end{figure}

The projection of the total angular momentum excluding rotation on the internuclear axis $\Omega$ is a conserved quantity, and we perform the calculation separately for each $\Omega$.The PECs for different values of $\Omega$ are in general degenerate. For $P$-wave contributions, this degeneracy is lifted by the spin-orbit interaction in the electron-neutral scattering state~\cite{deissObservationSpinorbitdependentElectron2020}. This effect is only important in heavy atoms, such as rubidium and cesium, and can be neglected for potassium.

\autoref{fig:allOmegaCs2} and \autoref{fig:allOmegaKCs} illustrate the effect of $\Omega$ on the PECs of selected states in Cs$_2$ and \KCs{}, respectively, where Cs is the perturber. At the position of the outermost minimum around $R=1400\,a_0$, the interaction is dominated by $S$-wave interactions and the PECs for different values of $\Omega$ remain almost degenerate. Moving towards shorter internuclear distances, we encounter a minimum in the $^{1,3}\Sigma^+$ PECs around $R=1250\,a_0$. Here the curves for different values of $\Omega$ have very different depths. We note that we have discussed these minima previously in the context of driving transitions to ion-pair states~\cite{peperFormationUltracoldIon2020a}. After solving the Schr\"odinger equation for the nuclear motion in the calculated $\Omega$-dependent PECs (see next section), we find that the binding energy of the lowest level in the outermost well varies by less than \SI{2}{\mega\hertz} (see wavefunctions and energies of the vibrational levels in \autoref{fig:allOmegaCs2} and \autoref{fig:allOmegaKCs}) for $\Omega<7/2$, and we thus report only the mean of these values.

We observe that the depth of the outermost well in the calculated PECs increases when (\emph{i}) the basis set includes more hydrogenic manifolds above the state of interest and (\emph{ii}) the reference state (Eq. 2 of the article) is changed from the higher-lying manifold to the low-$l$ asymptote of interest. The latter can be understood as follows: the binding at an internuclear separation close to the outer turning point of the Rydberg electron is dominated by $S$-wave scattering. When the quantum defect of the reference state is reduced, the semi-classical kinetic energy of the electron-atom scattering at a given distance also decreases (Eq. 2 of the article). Thus the value of the $^1S$ and $^3S$ scattering lengths decreases, \textit{i.e.}, the absolute magnitude of the (negative) $^3S$ scattering length increases and the well depth of the $^3\Sigma^+$ state increases. The binding of the mixed $^{1,3}\Sigma^+$ state is dominated by the interaction in the $^3S$ channel and thus also increases.

The interaction potentials correlating to the asymptotes \lrrm{K}{29}{P}{3/2}{K}{4}{S}{1/2}{1} and \ars{Cs}{31}{P}{3/2} \arshf{K}{4}{S}{1/2}{1} are presented in \autoref{fig:allOmegaK2} and \autoref{fig:allOmegaCsK}, respectively. The interaction potentials for K$_2$ and \CsK{} show an interesting change of the phase of the oscillatory potential at $R\approx\num{1200}\,a_0$. This is caused by the transition from $S$-wave scattering ($\propto|\Psi(R)|^2$) to $P$-wave scattering ($\propto|(\partial\Psi(R)/\partial R)|^2$) as the semi-classical energy of the Rydberg electron at the position of the perturber increases (see Eq. 2 of the article)~\cite{Omont1977}.

\begin{figure}[tb]
	\centering
	\includegraphics[width=\linewidth]{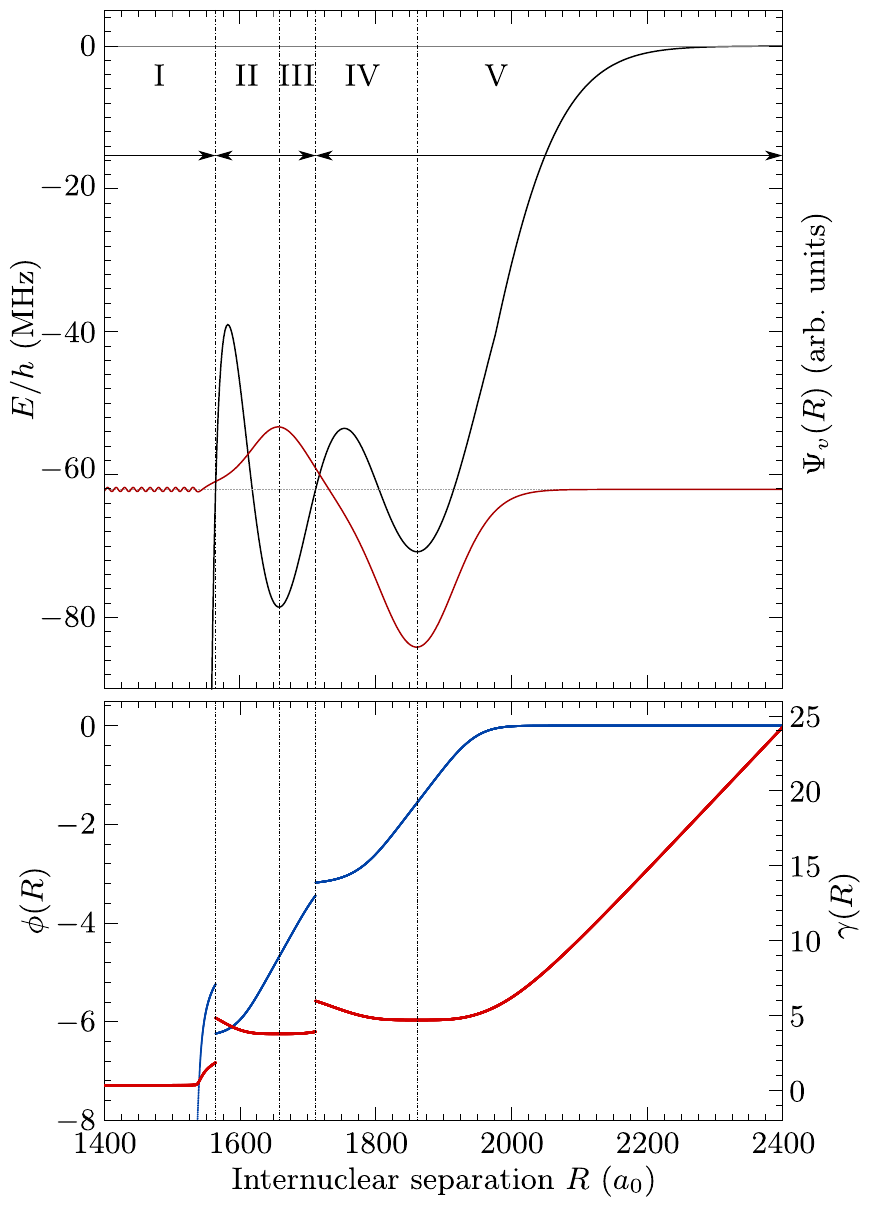}
	\caption{(Upper panel) Vibrational wave function $\Psi_v(R)$ (red line) of a bound state in the \lrrm{Cs}{35}{P}{j}{Cs}{6}{S}{1/2}{4} $^3\Sigma^+(\Omega=\nicefrac{1}{2})$ interaction potential. The integration was carried out separately in the region I to V in the directions indicated by the black arrows. (Lower panel) Obtained phase $\phi(R)$ (blue dots) and amplitude $\gamma(R)$ (red dots) after the matching procedure used to reconstruct the vibrational wave function $\Psi_v$ (red line in the upper panel).
		\label{fig:wavefunction}
	}
\end{figure}

\section{Calculation of vibrational wavefunctions}
\label{appendix:vibwfcts}

At small internuclear separations, the low-energy expansion of the Fermi-contact interaction breaks down, non-adiabatic couplings to other interaction potentials become important, and the behavior of the vibrational wavefunction is thus ill defined. Such a situation is, however, not uncommon in physics and several methods have been developed to overcome this problem, such as the stabilization method \cite{haziStabilizationMethodCalculating1970} and the Milne phase-amplitude method \cite{milneNumericalDeterminationCharacteristic1930,sidkyPhaseamplitudeMethodCalculating1999a}. For the calculation of vibrational wave functions, we adopt the modified Milne phase-amplitude method, as described by Sidky and Ben-Itzhak \cite{sidkyPhaseamplitudeMethodCalculating1999a}.

The central idea of the Milne method is that the wave function $\Psi_v(R)$ is replaced by the product of an $R$-dependent amplitude $\gamma(R)$ and the sine of an $R$-dependent phase $\phi(R)$
\begin{equation}
\label{eq:milnewavefunction}
	\Psi_v(R)=\sqrt{\frac{2\mu}{\uppi}}\exp (\gamma(R))\sin (\phi(R))
\end{equation}
in atomic units, with the reduced mass $\mu$ of the two-body problem. Making this substitution for the wavefucntion in the Schr\"odinger equation of nuclear motion yields a set of differential equations
\begin{equation}
\label{eq:milneone}
\difftwo{\gamma}{R}+\left(\diff{\gamma}{R}\right)^2-\left(\diff{\phi}{R}\right)^2+k^2(R)=0,
\end{equation}
\begin{equation}
\label{eq:milnetwo}
	\diff{\phi}{R}=\exp(-2\gamma).
\end{equation}

The advantage of Eqs.~\eqref{eq:milneone} and \eqref{eq:milnetwo} lies in the smooth behavior of the phase $\phi(R)$ and amplitude $\exp(\gamma(R))$ over the integration region, whereas $\Psi_v(R)$ is an oscillatory function. $\gamma(R)$ and $\phi(R)$ only remain smooth until the next maximum in the potential $V(R)$. Consequently the integration of Eqs.~\eqref{eq:milneone} and \eqref{eq:milnetwo} is conducted in separate regions, as illustrated in \autoref{fig:wavefunction} for the example of the \linebreak\lrrm{Cs}{35}{P}{3/2}{Cs}{6}{S}{1/2}{4} $^3\Sigma^+(\Omega=\nicefrac{1}{2})$ interaction potential. The numerical integration in each region is started at the minimum of the potential region with the amplitude given by the Wentzel–Kramers–Brillouin (WKB) approximation \cite{yooImplementationQuantumdefectTheory1986} and the starting phase of the integration is chosen as $\phi=0$. The wavefunction is then propagated to the next closest potential well, as indicated by the black arrows in the integration regions I to V indicated in the upper panel of \autoref{fig:wavefunction}.

\begin{figure}
	\centering
	\includegraphics[width=\linewidth]{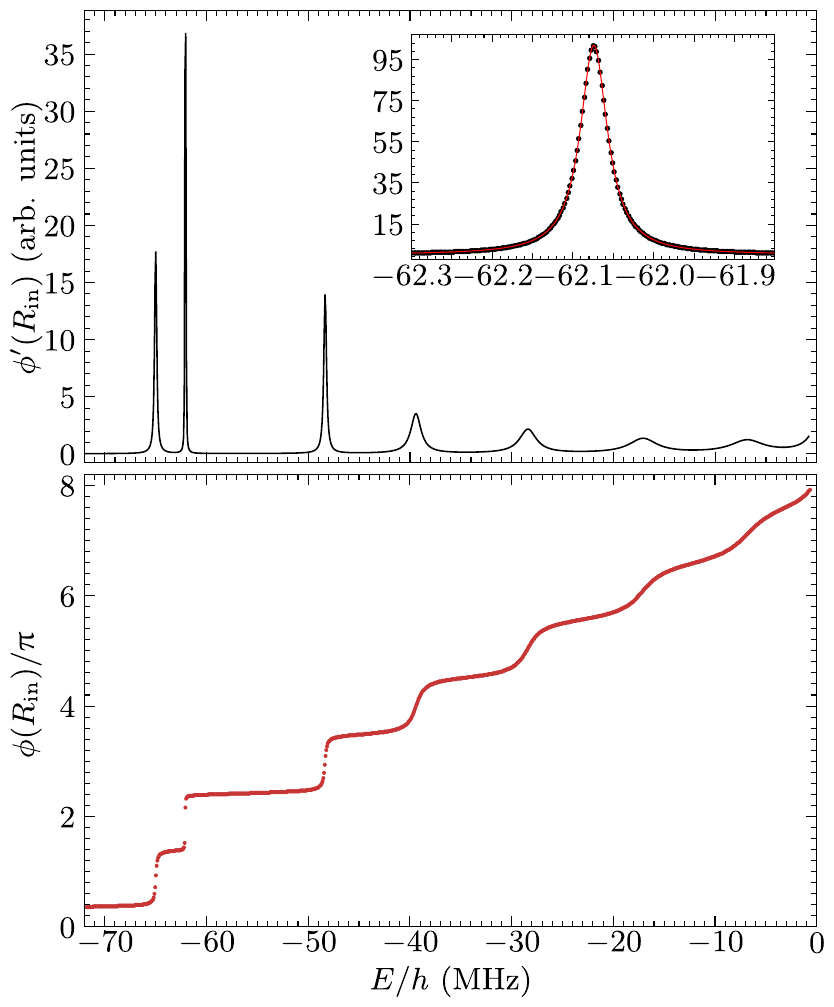}
	\caption{(Lower panel) Phase $\phi$ (red dots) of the wave function at an inner sampling point $R_\mathrm{in}$ at different energies $E$ below the \lrrm{Cs}{35}{P}{j}{Cs}{6}{S}{1/2}{4} in the interaction potential of the $^3\Sigma^+$ state. At the position of bound vibrational states, the value of the phase increases by $\uppi$. (Upper panel) Energy derivative of the phase $\phi'$ giving the positions and widths of bound vibrational states. The inset shows the behavior around the most narrow feature close to \SI{-62.1}{\mega\hertz} calculated using a narrower energy grid. The lifetime of the bound state can be extracted from the width of the calculated resonance by a Lorentzian line fit (red line).
		\label{fig:milnephaseandspectrum}
	}
\end{figure}

After the integration process, the solutions obtained for the separate integration regions are matched at the boundaries to obtain a continuous vibrational wave function $\Psi_v(R)$. This procedure relies on the fact that only the wavefunction $\Psi_v(R)$, and not the amplitude $\gamma(R)$ or phase $\phi(R)$, needs to be continuously differentiable. The phase in the outer classically forbidden region is kept at $\phi=0$ and the amplitude at an inner end point of the integration ($R_\mathrm{in}$) is kept at the value of the WKB approximation, ensuring energy normalization of the wave function. The resulting phase $\phi(R)$ and amplitude $\gamma(R)$ for a given energy $E$ are plotted in the lower panel of \autoref{fig:wavefunction}. The obtained wave function given by Eq.~\eqref{eq:milnewavefunction} is shown in red in the upper panel of \autoref{fig:wavefunction}. For a calculation of bound vibrational states, the phase at the inner end point of the integration $R_\mathrm{in}$ is plotted against the energy $E$ at which Eqs.~\eqref{eq:milneone} and \eqref{eq:milnetwo} are solved, as shown in \autoref{fig:milnephaseandspectrum}. When the energy of a bound vibrational state is reached, the phase jumps by $\uppi$, as can be seen in the lower panel of \autoref{fig:milnephaseandspectrum}. This property is an important advantage of the modified Milne phase-amplitude method over other methods such as the Numerov method, because the number of bound vibrational states in a certain region can be easily obtained by monitoring the respective phase shift and no bound states can be missed.

The spectrum of vibrational states can then be obtained from an energy derivative of the obtained phase $\left(\partial\phi(R_\mathrm{in},E)/\partial E\right)$, as shown in \autoref{fig:milnephaseandspectrum}. The position and lifetime of vibrational bound states are obtained by fitting a Breit-Wigner line profile. The width of the resonance features corresponds to the strength of the coupling $\Gamma_\mathrm{cpl}$ to the short-range quasi-continuum. The Milne method does not only yield the eigenenergies and wavefunctions of vibrational levels, but also the characteristic time $\tau_\mathrm{cpl}=\nicefrac{1}{\Gamma_\mathrm{cpl}}$ for which the molecules stay localized in the outermost wells of the electronic state \cite{wignerLowerLimitEnergy1955}. For example, for the resonance in the outermost potential well of the \lrrm{K}{29}{P}{3/2}{Cs}{6}{S}{1/2}{3} $^3\Sigma^+$ electronic state, we determine Breit-Wigner widths of $\Gamma_\mathrm{FWHM}=\SI{8.7}{\mega\hertz}$, $\Gamma_\mathrm{FWHM}=\SI{7.1}{\mega\hertz}$, and $\Gamma_\mathrm{FWHM}=\SI{6.1}{\mega\hertz}$ for  $\Omega=$1/2, 3/2, and 5/2, respectively~\cite{wignerLowerLimitEnergy1955}. Simulating the resulting spectrum with an equally weighted sum of Lorentzians, we extract a theoretical line width of \SI{7.2}{\mega\hertz}.  This is in good agreement with the experimental line width of \SI{7.7(7)}{\mega\hertz}. We interpret this width as natural line width of the resonance due to decay to shorter internuclear distances, where non-adiabatic transitions might cause a dissociation or autoionization of the long-range Rydberg molecule~\cite{butscher2011}.

We note that in a harmonic potential, the zero-point energy scales as $1/\sqrt{\mu_\mathrm{AB}}$ with the reduced mass $\mu_\mathrm{AB}$. The here calculated zero-point energies for the potentials with identical perturber follow this general trend, with significant deviations caused both by the anharmonicity of the PECs and the differences in the Born-Oppenheimer potentials.

\section{Assignment of photoassociation resonances}
\label{appendix:assignment}

The interpretation of the recorded spectra depends crucially on a correct assignment of the observed resonances to vibronic levels of the LRM. In the K$_2$ autoionization spectrum recorded below the \lrrm{K}{29}{P}{3/2}{K}{4}{S}{1/2}{1} asymptote, the assignment is rather unambiguous and we identify the two strong resonances above \SI{-100}{\mega\hertz} as the $v=0$ levels of the $^3\Sigma^+$ and $^{1,3}\Sigma^+$ states, as indicated by the solid and dashed magenta lines. The \CsK{} photoassociation spectrum recorded below the \lrrmnohf{Cs}{31}{P}{3/2}{K}{4}{S}{1/2} asymptote is more congested. We assign the $v=0$ levels of the $^3\Sigma^+$ and $^{1,3}\Sigma^+$ states to the resonances marked by the solid and dashed green lines because their binding energies are close to the calculated value, and the Cs PFI signals are much stronger than the \CsK{} autoionization signals, which is expected for the $v=0$ states in the outermost well (see Fig. 4 of the article). The $^3\Sigma^+$ resonance is accompanied by a second one at slightly larger detunings. For this more-strongly bound resonance, the relative yield of CsK$^+$ over Cs PFI ions is much larger than for the assigned resonance. We thus assume that this resonance originates from a state at shorter-range, which decays faster by autoionization. The resonances in the mm-wave spectrum below the \lrrm{Cs}{37}{S}{1/2}{K}{4}{S}{1/2}{2} asymptote are assigned based on the assignment and extrapolation of resonances at nearby asymptotes  \lrrm{Cs}{n}{S}{1/2}{K}{4}{S}{1/2}{2} (see previous section).

The assignment of the resonances in the \KCs{} spectrum recorded below the \lrrm{K}{29}{P}{3/2}{Cs}{6}{S}{1/2}{3} asymptote is again rather unambiguous. In the Cs$_2$ spectrum recorded below the \lrrm{Cs}{31}{P}{3/2}{Cs}{6}{S}{1/2}{3} asymptote, the $^3\Sigma^+$ $(v=0)$ resonance is easily identified. We assign the $^{1,3}\Sigma^+$ $(v=0)$ resonance to the peak at larger detunings based on the calculated resonance structure.

%

\end{document}